\documentclass[preprint,preprintnumbers,a4paper,11pt]{article}
\usepackage{latexsym}  
\usepackage{amssymb}
\usepackage{graphicx,color}
\usepackage{amsmath}

 \makeatletter
  \renewcommand{\theequation}{%
    \thesection.\arabic{equation}}
   \@addtoreset{equation}{section}
\newcommand{\eqn}[1]{&\hspace{-0.2em}#1\hspace{-0.2em}&}
\renewcommand\theequation{\arabic{section}.\arabic{equation}}
\begin{document}
\begin{titlepage}
\begin{center}
{\Large \bf Effective Valence Quark Model  and \\[.1in]
a Possible Dip in $d Br(B \to K \ell \bar{\ell})/dq^2$}
\vspace{4mm}

{\bf Hiroyuki Ishida$^a$ and Yoshio Koide$^b$}
\vskip 5mm
{${}^a$\it Department of Physics, Tohoku University, 
Sendai 980-8578, Japan\\
\it E-mail address: h\_ishida@tuhep.phys.tohoku.ac.jp\\
${}^b$\it Department of Physics, Osaka University, 
Toyonaka, Osaka 560-0043, Japan\\
\it E-mail address: koide@kuno-g.phys.sci.osaka-u.ac.jp}
\date{\today}

\vspace{3mm}

\begin{abstract} 
In rare $B$ meson decays $B\rightarrow K \ell^+ \ell^-$, 
a possible contribution of $\ell^+\ell^-$ emission via photon
from the ``spectator" quark $q$ ($q=u,d$) in the $B$ meson 
$(q\bar{b})$ is investigated in addition to the conventional one 
$\bar{b} \rightarrow \bar{s} +\gamma \rightarrow \bar{s}+\ell^+ + \ell^-$. 
If such a contribution is sizable compared with the standard 
estimation of $B\rightarrow K \ell^+ \ell^-$, we will observe
visible difference between  
$d \Gamma(B^0\rightarrow K^0 \ell^+ \ell^-)/d q^2$ and  
$d \Gamma(B^+\rightarrow K^+ \ell^+ \ell^-)/d q^2$ in 
$q^2$ dependence ($q^2 \equiv m_{\ell\ell}^2$).
Besides, as a result of the interference between the conventional 
one and a new one, a dip appears in 
$d \Gamma(B\rightarrow K \ell^+ \ell^-)/d q^2$
at a small region of $q^2$.  
The interference effect in the $B^0$ decay will also be observed 
differently from that in the $B^+$ decay. 
The calculation is done based on a 
semi-classical approach, a valence quark model. 
In the present model, the photon emission from the spectator 
quark $q$, $d \rightarrow d+\gamma$ ($u \rightarrow u+\gamma$)
is independent of the $b$-$s$ transition mechanism, and the 
characteristic results are due to a straightforward estimate 
of the quark propagator 
which cannot be incorporated into the factorization method. 
The model is not a valence quark ``dominant" model, 
so that, for example, the valence quarks 
in the final state carry 
only 24\% of the energy-momentum of the kaon. 
\end{abstract}

\end{center}
\end{titlepage}
%

\renewcommand \thesection{\arabic{section}}
\section{Introduction}

Recent observations of the bottom meson decays 
$B \rightarrow K \ell^+ \ell^-$ 
by Belle \cite{Belle} and BABAR \cite{Babar} seem to reveal 
an interesting feature: the observed $q^2$ dependence 
of the differential branching fraction,
$d Br( B \rightarrow K \ell^+ \ell^-)/d q^2$, 
seems to have a dip at a small value of $q^2$ 
($\equiv m_{\ell\ell}^2$),  
i.e. $q^2 \sim 1$ GeV$^2$. 
On the other hand, the LHCb experiments have reported a dip in 
$d Br( B^0 \rightarrow K^0 \mu^+ \mu^-)/d q^2$ \cite{LHCb2012dip} 
and no dip in $d Br( B^+ \rightarrow K^+ \mu^+ \mu^-)/d q^2$ 
\cite{LHCb2012non}. 
As we emphasize in the end of the final section,
these experimental results are very suggestive to us. 
Of course, we cannot deduce such the existence of a dip only
from the current $B$ decay data, because the amount of data is still 
not sufficient. 
Besides, we cannot see such a dip in the data of CDF \cite{CDF}. 
Nevertheless, in this paper, we dare to investigate a possibility 
that a dip in $d Br/dq^2$ is true, because it means that there 
is a new contribution to the decays $B \rightarrow K \ell^+ \ell^-$ in
addition to the conventional electroweak penguin decay \cite{B_theory},
\begin{eqnarray}
{\cal H}^{eff} = G^{eff}_{EW} \frac{1}{e} (\bar{s} 
\sigma_{\mu\nu} b_R) F^{\mu\nu} \,,\label{Eq:1.1}
\end{eqnarray}
where
\begin{eqnarray}
G_{EW}^{eff}=  \frac{G_F}{\sqrt{2}} \frac{\alpha}{\pi} 
V_{ts}^\ast V_{tb}\, 2 m_b\,,\label{Eq:1.2}
\end{eqnarray}
and, for simplicity, we have dropped contribution from $b_L$.
In the conventional analysis \cite{B_theory}, they use effective 
Hamiltonian to perform 
this transition (see for a review \cite{B_review}).
Although we have certainly $q^2$ dependence in their Hamiltonian,
we omit such term due to smallness of its Wilson coefficient.
As a result, the differential branching fraction does not have the $q^2$ pole 
and it cannot also explain the dip at small $q^2$ region.
On the other hand, in the recent analysis \cite{Feldmann:2002iw}-\cite{Jager:2012uw}, 
they have promoted to improve the analysis at the low recoil region, 
that is the large $\sqrt{q^2}$ of the order of the $b$-quark mass.

Actually, recent calculations about rare B meson decay channels 
seem to become high accuracy by developments of these studies.
In the research of flavor physics, however, it is essential to  
pay attention to flavor-dependent phenomena.
Any suggestions for flavor physics will not be 
obtained from flavor-blind phenomena. 
Unfortunately, recent development of the high 
energy physics and QCD seems to weaken characteristics
of the quark flavors. 
For example, recall that, during four years after discovery of 
the charmed mesons (1975), 
people had believed that lifetimes of the charmed mesons $D^+$ 
and $D^0$ are $\tau(D^+) \simeq \tau(D^0)$ speculated by QCD, until
a possibility of $\tau(D^+) \gg \tau(D^0)$ is pointed out in 1979
\cite{K-K_PRD79} 
and,  until 
the observation of  $\tau(D^+) > \tau(D^0)$ at SPEAR is, in fact, 
reported \cite{Luth79}.  
Again, let us direct our attention to $B \rightarrow K + \ell^+ + \ell^-$
decays.
There must be some differences between 
$B^+ \rightarrow K^+ \ell^+ \ell^-$ and 
$B^0 \rightarrow K^0 \ell^+ \ell^-$, as far as 
$\ell^+ \ell^-$ emission is done by photon, 
because of the charge difference between $q=u$ 
and $q=d$ in $B=(q\bar{b})$.
Nevertheless, most peoples have investigated 
only quantities without distinction between 
$B^+$ and $B^0$, because effects from the 
spectator quark $q$ seems to be negligibly small. 
In this paper, we would like again to pay attention
to valence quarks in hadrons without QCD corrections,
and thereby, we would like to find some differences 
among the quark flavors.  

The purpose of the present paper is not to discuss the 
absolute value of $Br( B \rightarrow K \ell^+ \ell^-)$
quantitatively, but to discuss the shape 
of $d Br( B \rightarrow K \ell^+ \ell^-)/d q^2$ qualitatively. 
We will speculate that if the ``observed" dip in the 
$q^2$ distribution of $B \rightarrow K \ell^+ \ell^-$ is 
true, a contribution due to photon emission from 
the ``spectator" quark\footnote{
The terminology ``spectator'' quark is somewhat misleading : 
In this case, the ``spectator'' quark means $q=u, d$ in 
the $B$ meson $(q\bar{b})$. 
In the conventional model, 
photon which produces a lepton pair is emitted via the effective interaction 
(\ref{Eq:1.1}), $\bar{b} \rightarrow \bar{s} +\gamma$. 
However, in the present paper, we discuss a case in which 
such photon is emitted from the ``spectator" quark 
$q=u, d$ as well as the $b\rightarrow s$ transition 
happens in the opposite side of the $q=u, d$. 
Nevertheless, we will use the terminology ``spectator" quark 
for $q=u, d$ in the $B$ meson $(q\bar{b})$ for convenience.
} 
is important. 
The first analysis of the ``spectator" quark contribution 
to $B \rightarrow K \ell^+ \ell^-$ has been done by 
Beneke, Feldmann and Seidel \cite{Beneke2001}.
(For a recent analysis, for example, see Ref.\cite{Khodjamirian2012} 
and the references therein.)
If contribution from the spectator quark to the 
$B \rightarrow K \ell^+ \ell^-$ is sizable, the $q^2$ dependence of 
$dBr(B \rightarrow K \ell^+ \ell^-)/d\, q^2$ will be considerably 
different between $B^0$ and $B^+$ 
in so far as there is a dynamics which can distinguish the spectator quarks.
Our interest is in this difference between $B^0$ and $B^+$ decays.  
(For isospin asymmetries, for example, see \cite{Feldmann:2002iw} and \cite{i-asym}.) 
In order to see the difference clearly, for the moment, we dare to drop 
form factor effects. 
For recent study of the form factor effects, for example, see 
Ref.\cite{form_factor}.
The purpose of the present paper is not to give a good fitting to 
the observed branching ratios and the differential branching ratios.
It is to make a comparison between photon emission from the 
spectator quark and that from $b\rightarrow s$ transition qualitatively.

\begin{figure}[!h]
\begin{center}
\includegraphics[width=8cm,clip]{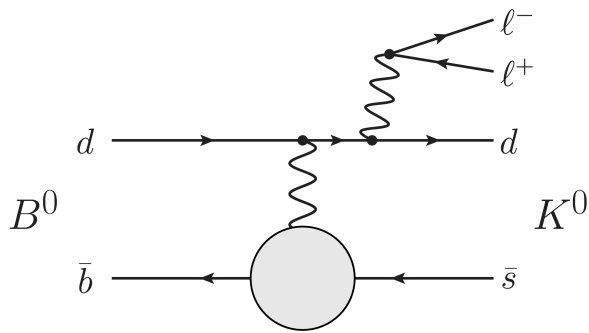}
\end{center}
\caption{Feynman diagram for 
$B^0 \rightarrow K^0 \ell^+ \ell^-$ due to 
photon emission from spectator quark.
}\label{Fig:1}
\end{figure}

Usually, the emission of photon from quarks is considered 
as that from the transition $b\rightarrow s$, Eq.~(\ref{Eq:1.1}), 
so that the decay amplitude has no $q^2$ pole. 
The interaction gives a decay amplitude of $B \rightarrow K \ell^+ \ell^-$
\begin{eqnarray}
{\cal M} = G^{eff}_{EW}  \frac{f_T(q^2)}{M_B +M_K} (P_B +P_K)^\mu 
[\bar{v}_\ell (k_2) \gamma_\mu u_\ell (k_1)] \,,
\label{Eq:1.3}
\end{eqnarray}
where $f_T(q^2)$ is a form factor in the meson currents 
for the effective quark interaction (\ref{Eq:1.1}). 
However, if the photon can be emitted from the 
``spectator'' quark line which can be seen in Fig.~\ref{Fig:1},
the decay amplitude will have a factor $1/q^2$ 
differently from the effective interaction (\ref{Eq:1.1}). 
In this paper, we consider a possibility that photon can 
be emitted from the ``spectator'' quark line, 
$d \rightarrow d\gamma$ as shown in Fig.~\ref{Fig:1}.
(Of course, we will take other three diagrams similar to 
Fig.~\ref{Fig:1} into consideration as discussed later.) 
The contribution (\ref{Eq:1.3}) is not entire one in the current 
estimates of the $B \rightarrow K \ell\ell$ decays.
There are actually many other contributions which are not included into this work.
For instance, the most well known contribution is so-called weak annihilation \cite{WA}.
We will use (\ref{Eq:1.3}) as the typical one of the 
conventional estimates. 
In our calculations, we do not apply any QCD 
corrections, {\it e.g.} form factor effects,
so that we will also neglect such corrections in the 
conventional contributions, too. 
Although such a treatment looks like an oversimplified one, 
 it is useful to see contributions from valence quarks individually.  
For example, we illustrate $q^2$ dependence in 
$dBr/d q^2$ later in Figs.~\ref{Fig:6} and \ref{Fig:8} by introducing 
a parameter $\xi$. 
Since we illustrate the standard model contribution 
by a curve with $\xi=0$, we can easily see 
corrected curves with $\xi \neq 0$ by imaging the
standard model contribution correctly for the curve 
with $\xi=0$. 
Of course, we do not consider that such neglected corrections 
are unnecessary.
Those considerations will become important in quantitative 
fitting to the data. 
However, in this paper, we give only qualitative study.  

%

The purpose of the present paper is to demonstrate sizable contribution of
 photon emission from the spectator quark rather than to propose a new 
mechanism of the $b$-$s$ transition.
In the present model, the photon emission from the spectator 
quark $d$ ($u$), $d \rightarrow d+\gamma$ ($u \rightarrow u+\gamma$) 
is independent of the $b$-$s$ transition mechanism, and 
the characteristic results are due to a straightforward estimate 
of the quark propagator which cannot be incorporated into 
the factorization method. 
Therefore, in the present paper, to specify the origin of the $b$-$s$ transition
is not essential.
At present, the most likely candidate of such a $b$-$s$ transition
will be the so-called gluon-penguin contribution 
without discrimination of spectator quarks. 
However, it is also interesting to consider another possibility,
an exchange of a family gauge boson $A_2^3$ 
as shown in Fig.~\ref{Fig:2} instead of the gluon penguin. 
Here $A_2^3$ is a family gauge boson which changes 
family number from ``2" to ``3".
We will give a brief review of the family gauge boson model 
\cite{K-Y_PLB_2012} in the next section (and also \ref{App:A}).
The results for $B^0$ and $B^+$ will be highly dependent as 
whether we adopt family gauge boson model or gluon penguin 
model. 

In Sec.~3, we discuss our assumptions in the effective 
valence quark model. 
In Sec.~4, we give a form factor-like function $f_+(q^2)$ 
which gives contribution of photon emission from quarks.  
(However, as we emphasize in Sec.~3, the factor $f_+(q^2)$ 
is not the so-called ``form factor".
In the present prescription, we do not introduce any 
form factor.
The factor $f_+(q^2)$ originates the existence of quark propagator 
seen in Fig.~\ref{Fig:1}.)
In Sec.~4, we put an assumption in order to calculate 
the function $f_+(q^2)$ simply.  
One of the purposes of the present paper is to demonstrate 
such $q^2$ dependence of the factors $f_+(q^2)$ given  
in Eqs.~(\ref{Eq:4.14}) - (\ref{Eq:4.17}) corresponding to four diagrams 
given in Fig.~\ref{Fig:3}.
The numerical results are given by Fig.~\ref{Fig:4} in Sec.~6. 
Our purpose is to see the individual contribution from each quark
to the photon emission as shown in Fig.~\ref{Fig:4} (a) - (d), so that 
the standard model contributions are oversimplified 
as given in Eq.~(\ref{Eq:1.3}) and we do not take QCD corrections
into consideration in this our naive results.  
Finally, Sec.~7 is devoted to the concluding remarks.  
Our results are somewhat different from the conventional one.
The reason of the difference is in that in the present calculation
we straightforwardly calculate effects of the quark propagator 
between the gauge-boson mediated vertex and the emitted photon vertex. 
We will emphasize the meaning of our prescription. 

%

\section{Another possibility of $b \rightarrow s$ transition}

\begin{figure}[!h]
\begin{center}
\includegraphics[width=10cm,clip]{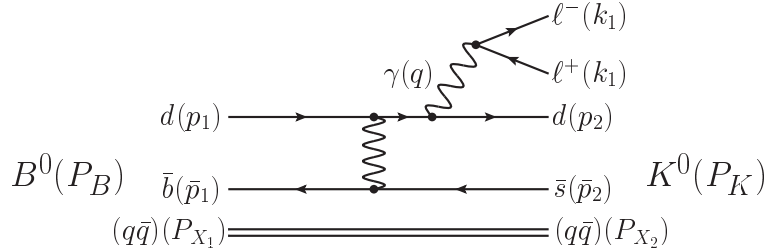}
\end{center}
\caption{Feynman diagram for $B^0 \rightarrow K^0 \ell^+ \ell^-$
and the momentum assignments.} \label{Fig:2}
\end{figure}

As we stated in Sec.~1, we demonstrate spectator effects 
based on a family gauge boson model.
As we show in Fig.~\ref{Fig:2}, the energy-momentum due to
$b$-$s$ transition is transmitted to the spectator quark $d$
mediated by a family gauge boson $A_2^3$.   

The family gauge boson model \cite{K-Y_PLB_2012} has 
somewhat peculiar characteristics differently from
conventional family gauge boson models.
The model has the following characteristics:  
(i) The family symmetry is U(3) [not SU(3)], so that
we have nine family gauge bosons (not eight those).  
(ii) The family gauge boson interactions 
are given by
\begin{eqnarray}
{\cal H}_{fam} = g_{F} \left[ (\bar{e}_i \gamma_\mu e_j) 
+ (\bar{\nu}_i \gamma_\mu \nu_j) 
+  U^{* d}_{ik} U_{jl}^d(\bar{d}_k \gamma_\mu d_l) 
+ U^{* u}_{ik} U_{jl}^u (\bar{u}_k \gamma_\mu u_l ) \right] (A_i^j)^\mu \,.\label{Eq:2.1}
\end{eqnarray}
Note that the family gauge boson mass 
matrix is diagonal on the basis in which the charged lepton mass 
matrix is diagonal, so that flavor-changing process 
appear only in the quark sector. 
(iii) $K^0$-$\bar{K}^0$, $D^0$-$\bar{D}^0$ and 
$B^0$-$\bar{B}^0$ mixings are caused only through  
non-zero quark-family mixing ($U^u \neq {\bf1}$ and 
$U^d \neq {\bf1}$). 
Note that if the U(3) family symmetry is broken by ${\bf 3}$ and/or 
${\bf 6}$ of U(3) as a conventional family gauge boson, 
a direct transition $A_i^j \leftrightarrow A_j^i$
will appear, while, in the present model, there are no such scalars. 
For example, $B_s^0$-$\bar{B}_s^0$ mixing is only caused through the 
down-quark mixing $U^d \neq {\bf 1}$.
If we suppose $U^d \simeq V_{CKM}$ (i.e. $U^u \simeq {\bf 1}$),
the gauge boson contribution to the $B_s^0$-$\bar{B}_s^0$ mixing is
suppressed enough by the CKM elements.\footnote{ 
Also note that the $P^0$-$\bar{P}^0$ mixing is mode with 
$\Delta_F=2$ ($N_F$ is family number), 
while $B \rightarrow K \ell \bar{ell}$
is mode with $\Delta_F=1$.  A kind of GIM mechanism \cite{GIM_PRD70} 
works only $\Delta_F=2$ mode in the quark sector \cite{harmless_13}.
}
(For more details, see \ref{App:A}.)

Such a family gauge boson model without a direct 
mixing $A_i^j \leftrightarrow A_j^i$ was first proposed by Sumino
\cite{Sumino_PLB09}. 
Therefore, the family gauge boson $A_2^3$ in the present model
cannot contribute to $B_s$-$\bar{B}_s$ mixing directly.  
Straightforwardly speaking, the mass of $A_2^3$ is 
independent of constraints from these 
ps-meson-anti-ps-meson mixings.
(iv) The gauge boson masses are given with
an inverted mass hierarchy 
i.e. $m^2(A_1^1), m^2(A_2^1), m^2(A_3^1) \gg 
m^2(A_2^2) \gg m^2(A_2^3) \gg m^2(A_3^3)$, 
so that we may suppose a mass of $A_2^3$ of an order of
$1 -10$ TeV \cite{YK_PRD_2013}. 

In the present paper, for the time being, we assume 
the mixing matrix among up-quarks is almost unit matrix, 
thus its mixings are negligibly small compared with that among down-quarks.
As a result, we discuss only the case of neutral B meson decay:
\begin{eqnarray}
B^0(P_B) \rightarrow K^0(P_K) + \ell^-(k_1) + \ell^+ (k_2) \,.\label{Eq:2.2}
\end{eqnarray}
We define momenta of quarks $\bar{b}$ and 
$d$ inside the bottom meson $B^0$ as $\bar{p}_1$ and $p_1$,
respectively, and $\bar{s}$ and $d$ inside $K^0$ as 
$\bar{p}_2$ and $p_2$, respectively as shown in Fig.~\ref{Fig:2}.
We also define the momentum of photon as $q$ in the decay 
$B^0(P_B) \rightarrow K^0(P_K) + \ell^-(k_1) +\ell^+(k_2)$, i.e.
\begin{eqnarray}
q \equiv k_1 + k_2 = P_B -P_K \,.\label{Eq:2.3}
\end{eqnarray}
Here, note that the momentum $P_B$ ($P_K$) is 
given by sum of momenta $p_1 + \bar{p}_1 + P_{X_1}$ 
($p_2 + \bar{p}_2 + P_{X_2}$) of the valence quarks and sea quarks:
\begin{eqnarray}
 \bar{p}_1 + p_1 +P_{X_1}= P_B\,,  \ \ \ \ 
 \bar{p}_2 + p_2 +P_{X_2} = P_K\,,\label{Eq:2.4}
\end{eqnarray}
where 
$P_{X_1}$ and $P_{X_2}$ are momenta of sea-quarks 
in the $B$ and $K$ mesons, respectively.

In order to know the momenta $\bar{p}_1$, $p_1$, $\bar{p}_2$
and $p_2$, we must reveal dynamical structures of the mesons. 
That is, in order to calculate the diagram given in Fig.~2,
we solve the dynamics of the system $(B \rightarrow K)$.
We integrate the diagram Fig.~2 with respect to the inner momenta
$p_1$, $p_2$, $\bar{p}_1$, $\bar{p}_2$, $P_{X1}$ and $P_{X2}$, 
and thereby, we can obtain the matrix element in terms of 
the observable quantities $P_B$ and $P_K$ only. 
In this paper, in an effort to calculate such new type diagram,
we propose an approach as a kind of the effective theory for 
valence quark diagrams.
In the next section, we represent those momenta $\bar{p}_1$, 
$p_1$, $\bar{p}_2$ and $p_2$ in terms of $P_B$ and $P_K$
with the help of an ``on-shell quark" assumption.
Thereby, we will estimate such diagrams given in Fig.~\ref{Fig:2}. 
Under this prescription, we will find that it is possible for photon
to be emitted from $d$ quark.

%

\section{Effective valence quark model}

In the present paper, we denote momenta of $B^0$, $K^0$, $\bar{b}$ and $d$ in
the $B^0$ meson, $s$ and $d$ in the neutral kaon $K^0$ as $P_B$, $P_K$, 
$\bar{p}_1$ and $p_1$, $\bar{p}_2$ and $p_2$, respectively.
Our assumption of ``on-shell quark" demands that quark masses are given by
\begin{eqnarray}
\bar{p}_1^2 = m_b^2\,, \ \ \ p_1^2 = m_{d1}^2\,, \ \ \ 
\bar{p}_2^2 = m_s^2\,, \ \ \ p_2^2 = m_{d2}^2 \,,\label{Eq:3.1}
\end{eqnarray}
where we have left a possibility that the mass of the $d$ quark in 
the bottom meson can be different from that of the $d$ quark in 
the kaon, so that we have denoted those as $m_{d1}$ and $m_{d2}$, 
respectively.
Here, it is our essential assumption that these quark masses are
almost constant for $q^2$, although those are still dependent 
on the energy scale $\mu$ of the system.

If we want to calculate a meson decay into a meson and something,
we must solve a composite state problem.
For example, in the $\bar{b}(x_b)$ and $d(x_d)$ system for the $B^0(X)$ meson,
two body bound state problem can be reduced into a one-body problem 
as to the relative coordinates $x=x_b -x_d$.
The variables $x=x_b -x_d$ and $X_B=(x_b+ x_d)/2$ corresponds to 
the momenta $p_b-p_d$ [($\bar{p}_1 -p_1)$ in the notation in Eq.~(\ref{Eq:3.1})] 
and $P_B=p_b +p_d$, respectively. 
However, in general, it is hard to solve such dynamics relativistically
and exactly.
Therefore, we usually use an easier method.
For example, we can treat the system as a two-body system of quark
and anti-quark system non-relativistically. 
Then, we must use effective quark masses (not the running 
quark masses $m_q(\mu)$) as masses of the constituents,
in which all of the effects of gluons, sea-quarks, and all the rest 
 are already taken into consideration.
 (For such a semi-classical approach to pseudo-scalar mesons, 
for example, see Ref.\cite{YK_PRD1981}.)     
Another easy method is to use the running quark mass values for 
the valence quarks, but is to consider that the valence quarks 
in the meson carry only a part of the momentum of the meson.
In this paper, we adopt the latter prescription. 

We define the fraction parameters $x_1$ and $x_2$
as follows:
\begin{eqnarray}
 \bar{p}_1 + p_1 = P_B -P_{X_1} \equiv x_1 P_B\,,  \ \ \ \ 
 \bar{p}_2 + p_2 =P_K -P_{X_2} \equiv x_2 P_K\,,\label{Eq:3.2}
\end{eqnarray}
where 
$x_1$ ($x_2$) is a fraction of momenta $p_1$ and $\bar{p}_1$ 
($p_2$ and $\bar{p}_2$) of the valence quarks $d$ and $\bar{b}$ 
($d$ and $\bar{s}$) versus the meson momentum $P_B$ ($P_K$). 
This is a big assumption, because Lorentz vector $\bar{p}_1 +p_1$
and $P_B$ cannot, in general, be connected each other by 
Lorentz scalar $x_1$.  
The parameters $x_1$ and $x_2$ are analogous to $x$ 
parameters in the high energy quark parton model in which 
$x$ distributions of the quark partons are well known 
(for a review, for example, see \cite{review_q-p}).
Usually, the matrix element of $B\rightarrow K$ is obtained 
by integrating with respected  to $x_i$ ($i=1,2$) over 
$x_{i min}\leq x_i \leq 1$.
However, in the present prescription, for simplicity, 
we will substitute special values $x_i(q^2_{max})$ 
for $x_i(q^2)$ 
as we show in Eqs.~(\ref{Eq:3.15}) and (\ref{Eq:3.16}) later. 

From the constraint (\ref{Eq:3.2}), we have the following relations
\begin{eqnarray}
x_1^2 M_B^2 +m_{d1}^2 -2 x_1 (p_1 P_B ) = m_b^2 \,, \ \ \ \ 
x_2^2 M_K^2 +m_{d2}^2 -2 x_2 (p_2 P_K ) = m_s^2 \,.\label{Eq:3.3}
\end{eqnarray}

Under the on-shell assumption, the quark momenta $p_1$ and $p_2$ 
can be expressed in terms of $P_B$ and $P_K$:
\begin{equation}
\begin{split}
p_1^\mu &= a_1 (P_B+ P_K)^\mu + b_1 (P_B- P_K)^\mu \,, \\
p_2^\mu &= a_2 (P_B+ P_K)^\mu + b_2 (P_K- P_B)^\mu \,,\label{Eq:3.4}
\end{split}
\end{equation}
where the coefficients $a_1$, $b_1$, $a_2$ and $b_2$ can, 
in general, be functions of $q^2$. 
[This is also a big assumption in this formulation. 
Note that we put this assumption only on momenta $(p_1,p_2)$,
but not on $(\bar{p}_1,\bar{p}_2)$ and $(P_{X1}, P_{X2})$, 
correspondingly to Fig.~\ref{Fig:2}.]
Then, we can obtain relations
\begin{eqnarray}
m_{d1}^2 \eqn{=} 
 a_1^2 [2(M_B^2 +M_K^2) -q^2] 
+ b_1^2\, q^2 + 2 a_1 b_1 \Delta_{BK}^2 \,,\label{Eq:3.5}\\
m_{d2}^2 \eqn{=} 
 a_2^2 [2(M_B^2 +M_K^2) -q^2] 
+ b_2^2\, q^2 - 2 a_2 b_2 \Delta_{BK}^2 \,,\label{Eq:3.6}
\end{eqnarray}
from Eq.~(\ref{Eq:3.1}), and 
\begin{eqnarray}
x_1^2 M_B^2 + m_{d1}^2 -m_b^2 \eqn{=} x_1 a_1 \left[ 
2(M_B^2 +M_K^2) + \Delta_{BK}^2 -q^2  \right]
+x_1 b_1 \left( \Delta_{BK}^2 + q^2 \right) 
 \,,\label{Eq:3.7}\\
%
x_2^2 M_K^2 + m_{d2}^2 -m_s^2 \eqn{=} x_2 a_2 \left[ 
2(M_B^2 +M_K^2) - \Delta_{BK}^2 -q^2 \right]
+ x_2 b_2 \left( -\Delta_{BK}^2 + q^2 \right) 
 \,,\label{Eq:3.8}
\end{eqnarray}
from Eq.~(\ref{Eq:3.4}),
where
\begin{eqnarray}
\Delta_{BK}^2 \equiv M_B^2 -M_K^2\,.\label{Eq:3.9}
\end{eqnarray}
Thus, if we give values of $x_1$ and $x_2$, 
we can completely determine the coefficients $(a_1, b_1)$ 
from the two relations (\ref{Eq:3.5}) and (\ref{Eq:3.7}), and  $(a_2, b_2)$ 
from the two relations (\ref{Eq:3.6}) and (\ref{Eq:3.8}), respectively.
Here, note that the replacement 
$(M_B, m_b, m_{d1}) \rightarrow (M_K, m_s, m_{d2})$ gives  
$(a_1, b_1) \rightarrow (a_2, b_2)$. 
Therefore, hereafter, we will discuss the relations only on 
$(a_1, b_1)$. 

The coefficients $(a_1, b_1)$ can be obtained as follows.
From Eq.~(\ref{Eq:3.5}), we obtain a relation between $a_1$ and $b_1$ 
(see \ref{App:D}): 
\begin{eqnarray}
b_1 = \frac{1}{q^2} \left[ - a_1 \Delta_{BK}^2 
\pm \sqrt{ D a_1^2 + m_{d1}^2 q^2} 
\right] \,,\label{Eq:3.10}
\end{eqnarray}
where
\begin{eqnarray}
D = \left[ (M_B -M_K)^2 -q^2\right] \left[ (M_B +M_K)^2 -q^2\right] 
\,.\label{Eq:3.11}
\end{eqnarray}
By substituting Eq.~(\ref{Eq:3.10}) into Eq.~(\ref{Eq:3.5}), we obtain a 
relation for $a_1$
\begin{eqnarray}
x_1^2 M_B^2 + m_{d1}^2 -m_b^2 = \frac{x_1}{q^2} 
\left[ - a_1 D 
\pm (\Delta_{BK}^2 +q^2 ) \sqrt{a_1^2 D + m_{d1}^2 q^2} 
\right] \,.\label{Eq:3.12}
\end{eqnarray}
The parameter $a_1$ can be obtained by solving Eq.~(\ref{Eq:3.12}) 
for $a_1$. 

The relation (\ref{Eq:3.12}) brings a new constraint to the model: 
Let us consider a limit of $q^2=q^2_{max}$, where 
\begin{eqnarray}
q^2_{max} \equiv (M_B -M_K)^2 \,,\label{Eq:3.13}
\end{eqnarray}
and it gives
\begin{eqnarray}
D(q^2) |_{q^2= q^2_{max}} = 0 \,.\label{Eq:3.14}
\end{eqnarray}
Therefore, the relation (\ref{Eq:3.12}) at a limit of $q^2=q^2_{max}$
leads to a constraint
\begin{eqnarray}
x_1 M_B = m_b \pm m_{d1} \,.\label{Eq:3.15} 
\end{eqnarray}
Note that the parameter $x_1$ in the definition (\ref{Eq:3.2}) was dependent
on the inner momentum $P_{X1}$ (i.e. $\bar{p}_1+p_1$), while 
$x_1$ given in (\ref{Eq:3.15}) is a constant (although the value $x_1$ given 
in (\ref{Eq:3.15}) is still dependent on the energy scale $\mu$). 
The crucial assumption is that we can approximately use 
the value of $x_1(q^2_{max})$ instead of $x_1(q^2)$ for whole physical 
region $q^2_{min} \leq q^2 \leq q^2_{max}$. 

Similarly, we obtain a constraint 
\begin{eqnarray}
x_2 M_K = m_s \pm m_{d2} \,.\label{Eq:3.16}
\end{eqnarray}
Note that the sign $\pm$ in Eq.~(\ref{Eq:3.15}) corresponds to the
sign $\pm$ in the relation (\ref{Eq:3.12}), but the sign
$\pm$ in Eq.~(\ref{Eq:3.15}) and $\pm$ in Eq.~(\ref{Eq:3.16}) are independent
each other. 

Quark masses $m_b$, $m_s$ and $m_d$ are function of 
the energy scale $\mu$, but it does not mean that 
those are always function of $q^2$ directly.
We consider that the quark mass values $m_b$, $m_s$ and $m_d$ 
in the $B \rightarrow K \ell^+ \ell^-$  decays are described 
by those at $\mu \sim M_B$. 
Then, we can estimate the values of $x_1$ and $x_2$ from 
Eqs.~(\ref{Eq:3.15}) and (\ref{Eq:3.16}), so that we can also estimate 
the values of $(a_1, b_1)$ and $(a_2, b_2)$.

More discussions from a phenomenological point of view  
will be given in Sec.~4.

%
\section{$\ell^+ \ell^-$ emission from valence quark:
$q \rightarrow q +\gamma$ vs $\bar{b} \rightarrow \bar{s} +\gamma$
in $B=(q\bar{b})$ }

\begin{figure}[h]
\begin{center}
\begin{minipage}{7cm}
\includegraphics[width=7cm,clip]{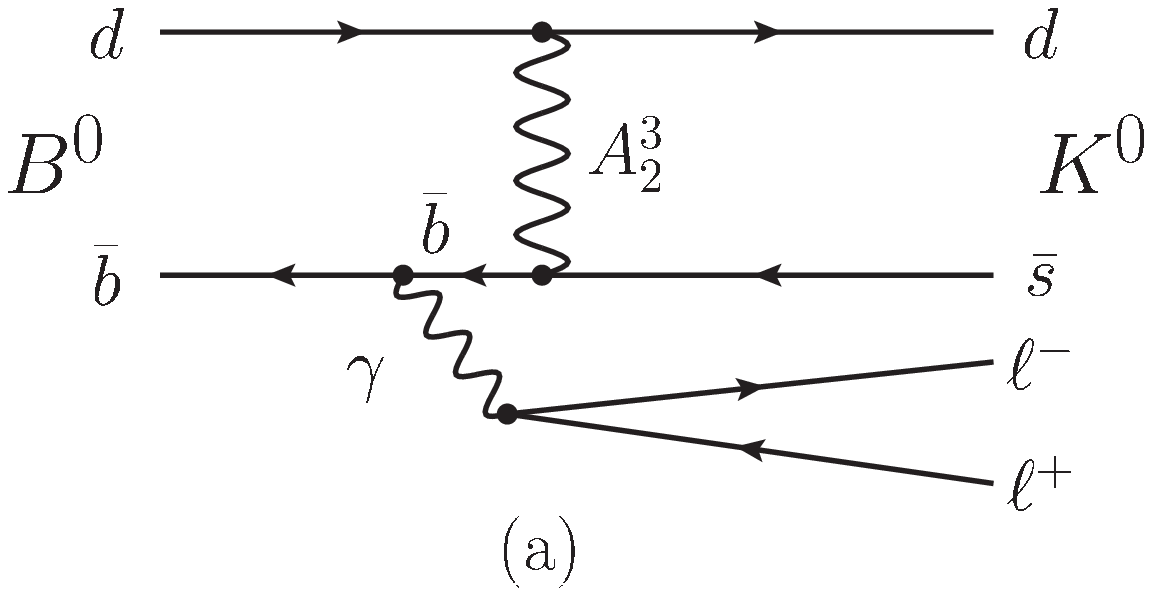}
\end{minipage}~~~~~
\begin{minipage}{7cm}
\includegraphics[width=7cm,clip]{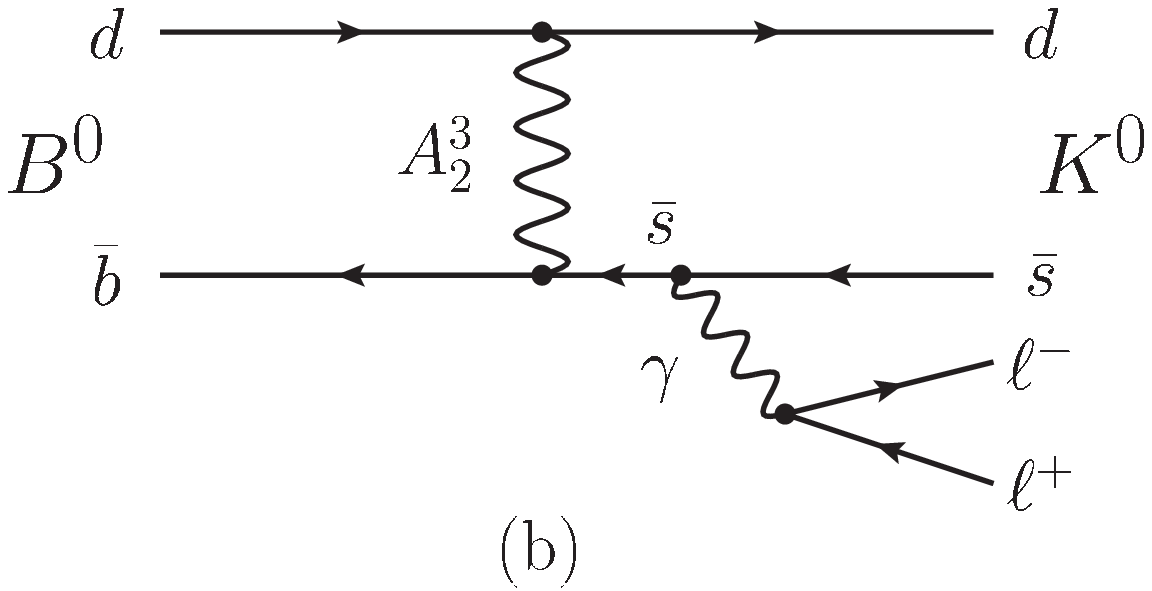}
\end{minipage}\\
\vspace{5mm}

\begin{minipage}{7cm}
\includegraphics[width=7cm,clip]{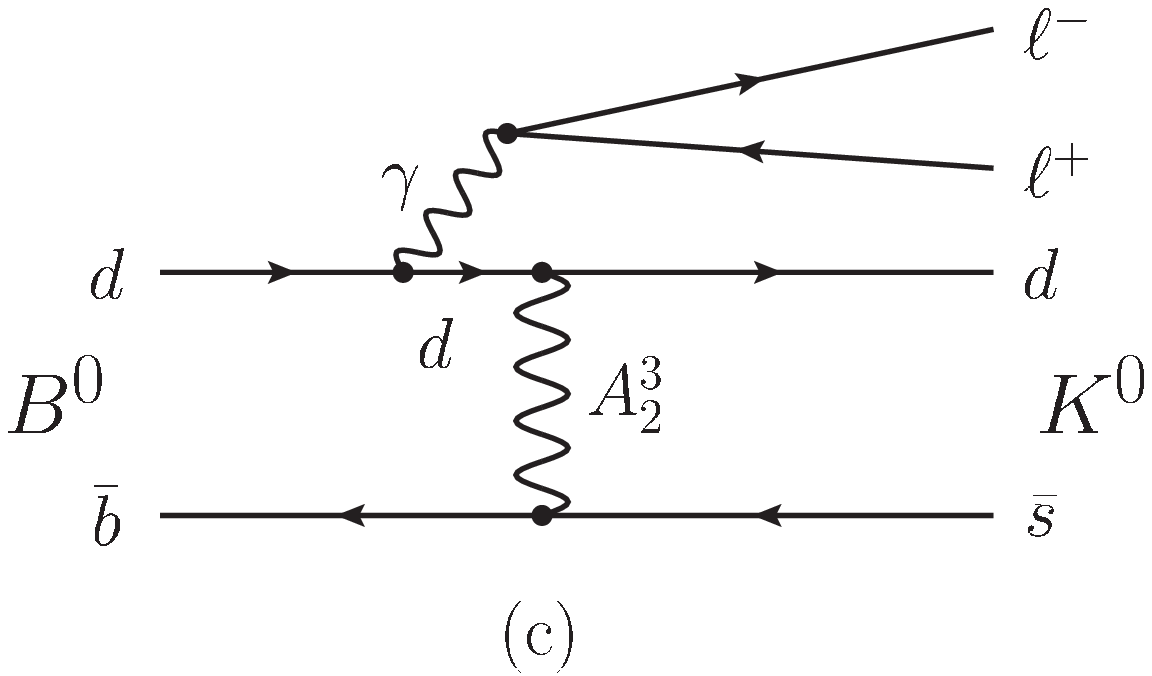}
\end{minipage}~~~~~
\begin{minipage}{7cm}
\includegraphics[width=7cm,clip]{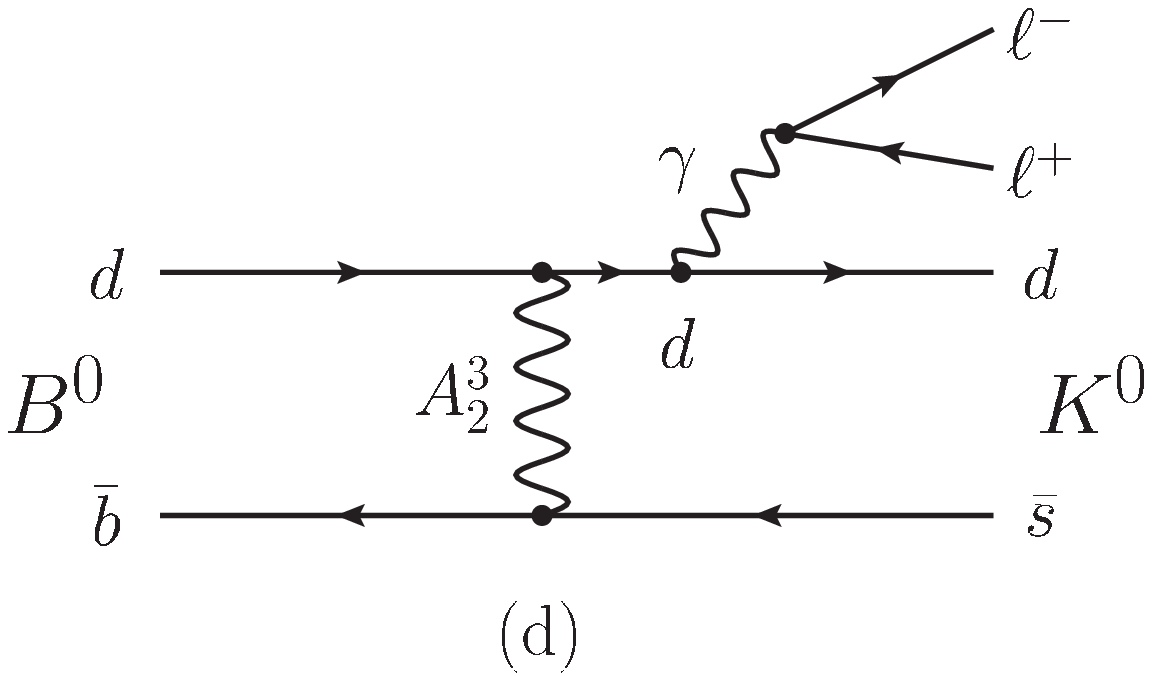}
\end{minipage}
\end{center}
\caption{ Feynman diagrams for $B^0 \rightarrow K^0 \ell^+ \ell^-$. } \label{Fig:3}
\end{figure}

We assume the following interactions for 
$B \rightarrow K \ell^+ \ell^-$ in addition to the conventional
$b \rightarrow s +\gamma$ interaction (\ref{Eq:1.1}):
\begin{eqnarray}
{\cal H} = \sum_{q=u,d,b,s} e_q (\bar{q} \gamma_\mu q) A^\mu 
- \sum_{\ell=e,\mu} e (\bar{\ell} \gamma_\mu \ell) A^\mu
+ \sum_{q= u,d} G_{fam}^q (\bar{b} \gamma_\rho s) (\bar{q} \gamma^\rho q)\,,\label{Eq:4.1}
\end{eqnarray}
where $e_d=e_s=e_b =-e/3$, $e_u=2e/3$, and 
\begin{eqnarray}
G_{fam}^q = \frac{g_F^2}{M_{23}^2} 
U_{33}^{\ast d} U_{22}^d U_{21}^{\ast q} U_{31}^q \,.\label{Eq:4.2}
\end{eqnarray}
Here, $U_{ij}^q$ are mixing matrix elements among quarks 
$q=(q_1, q_2,q_3)$, and
$M_{23}$ is a mass of a family gauge boson $A_2^3$. 
Based on the interactions (\ref{Eq:4.1}), we calculate the 
following four diagrams for 
$B^0 \rightarrow K^0 \ell^+ \ell^-$ as shown in 
Fig.~\ref{Fig:3}.\footnote{
We may consider that the contributions given 
in  Fig.~\ref{Fig:3} (a) - (b) are already included in the 
standard model contributions for the case of 
the gluon-penguin instead of the $A_2^3$ exchange. 
However, we go on this prescription in order to
see effects of photon emission from non-spectator quark.
}
Calculations corresponding to Fig.~\ref{Fig:3} (a) - (d) based on the 
standard model have already been given, for example, 
in Ref.\cite{Matsumori-Sanda_PRD}.
In our prescription, especially, a role of the propagator 
in Fig.~\ref{Fig:3} (quark-line which is not connected directly to
the mesons $B$ and $K$) is investigated.   

Hereafter, since we are interested in a case 
$B^0 \rightarrow K^0 \ell^+ \ell^-$, 
we will calculate only the case. 
Another case $B^+ \rightarrow K^+ \ell^+ \ell^-$
can easily be obtained by replacing $e_d \rightarrow e_u$
and $U^d \rightarrow U^u$.

Denominators of the propagators with momenta $\ell$ shown in 
Figs.~\ref{Fig:3} (a), (b), (c) and (d) are given as follows:
\begin{eqnarray}
\Delta_a \eqn{\equiv} \ell_{(a)}^2 -m_b^2  = q^2 -2 \bar{p}_1 q \,,\label{Eq:4.3}\\
\Delta_b \eqn{\equiv} \ell_{(b)}^2 -m_s^2 = q^2 +2 \bar{p}_2 q \,,\label{Eq:4.4}\\
\Delta_c \eqn{\equiv} \ell_{(c)}^2 -m_{d1}^2 =  q^2 -2 {p}_1 q \,,\label{Eq:4.5}\\
\Delta_d \eqn{\equiv} \ell_{(d)}^2 -m_{d2}^2 =  q^2 +2 {p}_2 q\,.\label{Eq:4.6}
\end{eqnarray}
By using the coefficients defined by Eq.~(\ref{Eq:3.4}), the expressions 
(\ref{Eq:4.3}) - (\ref{Eq:4.6}) are rewritten as follows:
\begin{eqnarray}
\Delta_a \eqn{=} (2 a_1 -x_1) \Delta_{BK}^2 + (1-x_1 +2 b_1) q^2\,,\label{Eq:4.7}\\
\Delta_b \eqn{=} -(2 a_2 -x_2) \Delta_{BK}^2 + (1-x_2 +2 b_2) q^2\,,\label{Eq:4.8}\\
\Delta_c \eqn{=} - 2a_1 \Delta_{BK}^2  + (1 -2 b_1) q^2\,,\label{Eq:4.9}\\
\Delta_d \eqn{=}  2a_2 \Delta_{BK}^2  + (1 -2 b_2) q^2 \,.\label{Eq:4.10}
\end{eqnarray}

In order to translate effective interactions among quarks
into hadronic fields, we use 
\begin{eqnarray}
\langle 0| (\bar{b}\gamma^\mu \gamma_5 d) |B^0(P_B)\rangle
= -i P^\mu_B f_B \,,\label{Eq:4.11}
\end{eqnarray}
for the initial state and apply same approach to the final state. 
The details to obtain amplitudes which correspond to
the diagrams (a), (b), (c) and (d) in Fig.~\ref{Fig:3} are
given in \ref{App:B}.

When we use the expression (\ref{Eq:3.4}), we obtain the following
form for the meson currents: 
\begin{eqnarray}
\hspace{-7mm}
{\cal M} = i \frac{1}{6} \bar{e}_b e f_K f_B G_{fam} \frac{1}{2}
\left[ f_+ (q^2) (P_B +P_K)_\mu + f_- (q^2) (P_B -P_K)_\mu \right]
\frac{1}{q^2} [\bar{v}_\ell(k_2) \gamma^\mu u_\ell(k_1)]\,,\label{Eq:4.12}
\end{eqnarray}
where $G_{fam}$ is defined by Eq.~(\ref{Eq:4.2}) and we have dropped 
the index $q=d$ because it is obvious that we calculate a 
case of $B^0 \rightarrow K^0 \ell^+ \ell^-$.

The second term with $q_\mu =(P_B -P_K)_\mu$ in Eq.~(\ref{Eq:4.12}) does not 
contribute the decay amplitudes because of
$q_\mu [\bar{v}_\ell(k_2) \gamma^\mu u_\ell(k_1)] =0$ for 
$m_{\ell 1} = m_{\ell 2}$.
For the expression $f_+(q^2)$, we obtain
\begin{eqnarray}
f_+(q^2) = f_{+}^a(q^2) + f_{+}^b(q^2) - f_{+}^c(q^2) - f_{+}^d(q^2) \,,\label{Eq:4.13}
\end{eqnarray}
where
\begin{eqnarray}
f_{+}^a(q^2) \eqn{=} \frac{ (x_1-2 a_1)M_K^2 +(1-x_1+ a_1 +b_1) q^2 }{
-(x_1-2 a_1) \Delta_{BK}^2 +(1-x_1+ 2 b_1 )q^2} \,,\label{Eq:4.14}\\
f_{+}^b(q^2) \eqn{=} \frac{ (x_2-2 a_2) M_B^2 +(1-x_2+ a_2 +b_2) q^2 }{
(x_2-2 a_2) \Delta_{BK}^2 +(1-x_2+ 2 b_2) q^2} \,,\label{Eq:4.15}\\
f_{+}^c(q^2) \eqn{=} \frac{ 2 a_1 M_K^2 + ( 1-a_1 -b_1) q^2 }{ 
 - 2a_1 \Delta_{BK}^2  + (1 -2 b_1) q^2} \,,\label{Eq:4.16}\\
f_{+}^d(q^2) \eqn{=} \frac{ 2 a_2 M_B^2 + (1- a_2 -b_2) q^2}{
2a_2 \Delta_{BK}^2  + (1 -2 b_2) q^2} \,.\label{Eq:4.17}
\end{eqnarray}
Note that these factors $f_+(q^2)$ given in Eqs.~(\ref{Eq:4.14}) - (\ref{Eq:4.17})
are not the so-called ``form factor" which denotes a quark 
structure. 
The functions, $f_+^a(q^2)$, $f_+^b(q^2)$, $f_+^c(q^2)$ and $f_+^d(q^2)$, 
originate in the propagators shown in Fig.~\ref{Fig:3} (a) - (d). 
Although we do not introduce any form factor since it 
is not a main story in the present prescription, 
this does not mean that we deny the existence of such 
form factors. 
It will become important to take such effects into consideration 
in an extended study in a future.    

So far we have discussed the case of the decay mode 
$B^0 \rightarrow K^0 \ell^+ \ell^-$, because we have
considered that up-quark mixing will be considerably 
small compared with down-quark mixing,  
$|U^u_{ij}|^2 \ll |U^d_{ij}|^2$. 
However, we can easily calculate the case 
$B^+ \rightarrow K^+ \ell^+ \ell^-$ 
similarly to the case $B^0 \rightarrow K^0 \ell^+ \ell^-$:  
A form of $f_+(q^2)$ for the decay $B^+ \rightarrow K^+ \ell^+ \ell^-$ 
can be obtained by replacing $e_d=-e/3 \rightarrow e_u=+2e/3$ in 
Eq.~(\ref{Eq:4.13}), i.e. 
\begin{eqnarray}
f_+(q^2) = f_{+}^a(q^2) +f_{+}^b(q^2) +2 f_{+}^c(q^2) +2 f_{+}^d(q^2) \,.\label{Eq:4.18}
\end{eqnarray}

%

\section{Interference effect in $d \Gamma/d q^2$} 

The partial decay width $\Gamma (B \rightarrow K \ell^+ \ell^-)$
is calculated from the matrix element 
\begin{eqnarray}
{\cal M} = G \left( 1 + \xi \frac{f_+(q^2)}{q^2} \right) 
(P_B +P_K)_\mu [\bar{v}_\ell(k_2) \gamma^\mu u_\ell (k_1) ] \,,\label{Eq:5.1}
\end{eqnarray}
where 
\begin{eqnarray}
G = G^{eff}_{EW} \frac{2 m_b f_T(0)}{M_B + M_K} \,.\label{Eq:5.2}
\end{eqnarray}
Here, for simplicity, we have neglected the $q^2$ dependence of the 
form factor $f_T(q^2)$ in the conventional model. 
(The numerical results are not almost change even if we take
the $q^2$ dependence of $f_T(q^2)$ into consideration. 
We will demonstrate it in \ref{App:E}.)  
The parameter $\xi$ is defined by
\begin{eqnarray}
\xi =  \frac{g_{fam}^2}{g_w^2} \frac{8 M_w^2}{M_{23}^2} 
\frac{U_{33}^{\ast d} U_{22}^d U_{21}^{\ast d} U_{31}^d }{V_{ts}^* V_{tb} } 
\frac{\pi^2}{9} 
 \frac{M_B + M_K}{2 m_b f_T(0)} f_K f_B \,.\label{Eq:5.3}
\end{eqnarray}
Certainly, this parameter should be $\left| \xi \right| \sim 10^{-5} {\rm GeV}^2$ 
with $M_{23} \sim$ a few TeV at a rough estimation 
in the family gauge boson model.
But, at present, we regard this parameter $\xi$ as a free parameter 
whose value is phenomenologically 
determined by the observed $q^2$ dependence of $dBr/dq^2$. 
Let us define a function $F(q^2)$ as
\begin{eqnarray}
G^2 F(q^2) \equiv \left. \frac{d \Gamma}{d q^2} \right|_{\xi=0}
 = \frac{1}{(2\pi)^3} \frac{1}{32 M_B^3} 
\int_{y_1}^{y_2} dy\, |{\cal M}|^2_{\xi=0} \,,\label{Eq:5.4}
\end{eqnarray}
where $ y \equiv m^2_{\ell K} =(k_2 +P_K)^2$, and
$y_1 = y_{min}$, $y_2=y_{max}$.
Then, $d \Gamma/d q^2$ is given by
\begin{eqnarray}
\frac{d\Gamma}{d q^2} (B \rightarrow K \ell^+ \ell^-) =  
G^2 \left( 1 + \xi \frac{f_+(q^2)}{q^2} \right)^2 F(q^2) \,.\label{Eq:5.5}
\end{eqnarray}
The explicit form of $F(q^2)$ is appeared in \ref{App:C}. 

Now, we can numerically evaluate the function $f_+(q^2)$ 
and ${d\Gamma}/{d q^2}$ by using these formulas 
(\ref{Eq:5.1}) - (\ref{Eq:5.5}).
First, we give quark mass values $m_b(\mu)$, $m_s(\mu)$,
and $m_{d1}(\mu)=m_{d2}(\mu)$ at $\mu =M_B -M_K$.
Then, we obtain the values, $x_1$ and $x_2$, 
by the relations (\ref{Eq:3.15}) and (\ref{Eq:3.16}). 
We assume that the quark mass values in this prescription
are almost independent of $q^2$, and those are only 
dependent on the value $\mu$.
We assume that these quark masses at $\mu =M_B -M_K$
are approximately not so deviated from those at $\mu =M_B$, 
so that we use the values which are determined by using
(\ref{Eq:3.15}) and (\ref{Eq:3.16}).
The coefficients $a_1$ ($a_2$) can be 
obtained by using Eq.~(\ref{Eq:3.12}) and then $a_1$ ($a_2$) 
can be gotten by using Eq.~(\ref{Eq:3.10}). 
We will obtain two solutions for Eq.~(\ref{Eq:3.12}). 
Note that the coefficients $(a_1, b_1)$ and $(a_2, b_2)$
are, in general, given as functions of $q^2$.

However, in order to give a more concise form of 
$(a_1, b_1)$ and $(a_2, b_2)$,  
let us put the following assumption from  phenomenological point of view:
These coefficients have no $q^2$ dependence approximately.
This demands $a_1=b_1$ ($a_2=b_2$) as 
seen in Eqs.~(\ref{Eq:3.5}) and (\ref{Eq:3.7}) [Eqs.~(\ref{Eq:3.6}) and (\ref{Eq:3.8})].
Then, we obtain concise forms
\begin{eqnarray}
a_1 = b_1 = \pm \frac{m_{d1}}{2 M_B} \,, \ \ \ 
a_2 = b_2 = \pm \frac{m_{d2}}{2 M_K} \,, \label{Eq:5.6}
\end{eqnarray}
from Eq.~(\ref{Eq:3.12}).
The sign $\pm$ in (\ref{Eq:5.6}) corresponds to $\pm$ in Eq.~(\ref{Eq:3.12}),
but the sign $\pm$ in $a_1=b_1$ need not to correspond 
to that in $a_2=b_2$. 
By using these solutions in Eq.~(\ref{Eq:5.6}), the expressions (\ref{Eq:4.14}) - (\ref{Eq:4.17}) 
are rewritten as follows:
\begin{eqnarray}
f_{+}^a(q^2) \eqn{=} \frac{ m_b M_K^2 +(M_B -m_b) q^2 }{
- m_b \Delta_{BK}^2 +(M_B -m_b)q^2} \,,\label{Eq:5.7}\\
f_{+}^b(q^2) \eqn{=} \frac{ m_s M_B^2 +(M_K -m_s) q^2 }{
 m_s \Delta_{BK}^2 +(M_K -m_s) q^2} \,,\label{Eq:5.8}\\
f_{+}^c(q^2) \eqn{=} \frac{ \pm m_{d1} M_K^2 + ( M_B \mp m_{d1}) q^2 }{ 
 \mp m_{d1} \Delta_{BK}^2  + (M_B \mp m_{d1}) q^2} \,,\label{Eq:5.9}\\
f_{+}^d(q^2) \eqn{=} \frac{ \pm m_{d2} M_B^2 + (M_K \mp m_{d2}) q^2}{
\pm m_{d2} \Delta_{BK}^2  + (M_K \mp m_{d2}) q^2} \,.\label{Eq:5.10}
\end{eqnarray}
Note that $f_{+}^a(q^2)$ and $f_{+}^b(q^2)$ are independent of
the choices $\pm$ in Eq.~(\ref{Eq:5.6}), but $f_{+}^c(q^2)$ and $f_{+}^d(q^2)$ 
are dependent on the choices.
If we take the positive sign for $a_1=b_1$ in (\ref{Eq:5.6}), then the function
$f_{+}^c(q^2)$ will have a pole at 
$q^2=m_{d1} \Delta_{BK}^2 /(M_B -m_{d1})$.
Also, if we take the negative sign in Eq.~(\ref{Eq:5.6}), then the function
$f_{+}^d(q^2)$ will have a pole at 
$q^2=m_{d2} \Delta_{BK}^2 /(M_K +m_{d2})$.
Therefore, in the numerical estimate of $d \Gamma/d q^2$, 
we take the signs in Eq.~(\ref{Eq:5.6}) as follows:
\begin{eqnarray}
a_1 = b_1 = - \frac{m_{d1}}{2 M_B} \,, \ \ \ 
a_2 = b_2 = + \frac{m_{d2}}{2 M_K} \,. \label{Eq:5.11}
\end{eqnarray}
Then, the propagator effects at $q^2=0$ are given by
\begin{eqnarray}
f_+^a(0) =f_+^c(0) = - \frac{M_K^2}{M_B^2 -M_K^2}\,, \ \ \ 
f_+^b(0) =f_+^d(0) = + \frac{M_B^2}{M_B^2 -M_K^2}\,. \label{Eq:5.12}
\end{eqnarray} 
 
%

\section{Numerical results}

For numerical estimates, for convenience, we adopt 
quark mass values \cite{q-mass} at $\mu = m(m_b) = 4.34$ GeV 
in place of those at $\mu =M_B-M_K$:
\begin{eqnarray}
m_b=4.34 \ {\rm GeV}\,, \ \ \ m_s=0.127 \ {\rm GeV}\,, \ \ \  
m_d \equiv m_{d1} = m_{d2} =0.00637 \ {\rm  GeV}\,.\label{Eq:6.1}
\end{eqnarray}
The input values (\ref{Eq:6.1}) lead to the following values of the
fraction factors $x_1$ and $x_2$:
\begin{eqnarray}
x_1(B) = 0.821, \ \ \ \ x_2(K)=0.244 \,,\label{Eq:6.2}
\end{eqnarray}
from Eqs.~(\ref{Eq:3.15}) and (\ref{Eq:3.16}), respectively.
We may have another choice.
However, numerical results are almost similar. 
Hereafter, we use the values (\ref{Eq:6.1}) as typical values in our 
prescription.

The values (\ref{Eq:6.2}) mean that the valence quarks $b$ and $d$ 
are almost dominant in the $B$ meson, while the valence quarks 
$s$ and $d$ carry only 24\% of the momentum of the kaon in 
the final state.
However, the value $x_2(K)=0.244$ is not common in the
all kaon processes, but the value $x_2(K)=0.244$ is one only 
in the case of $B \rightarrow K \ell^+ \ell^-$. 
For example, in a kaon decay (note that the value is not 
$x_2$, but it is $x_1$ because $K$ is one in the initial
state), we will again obtain a value close to
$x_1(K) \simeq 1$, because in this times we will use quark mass 
values $m_s(\mu)$ and $m_d(\mu)$ at $\mu= M_K$ (not $\mu=M_B$).

\begin{figure}[!h]
\begin{center}
\begin{picture}(90,125)(70,0)
  \includegraphics[height=.195\textheight]{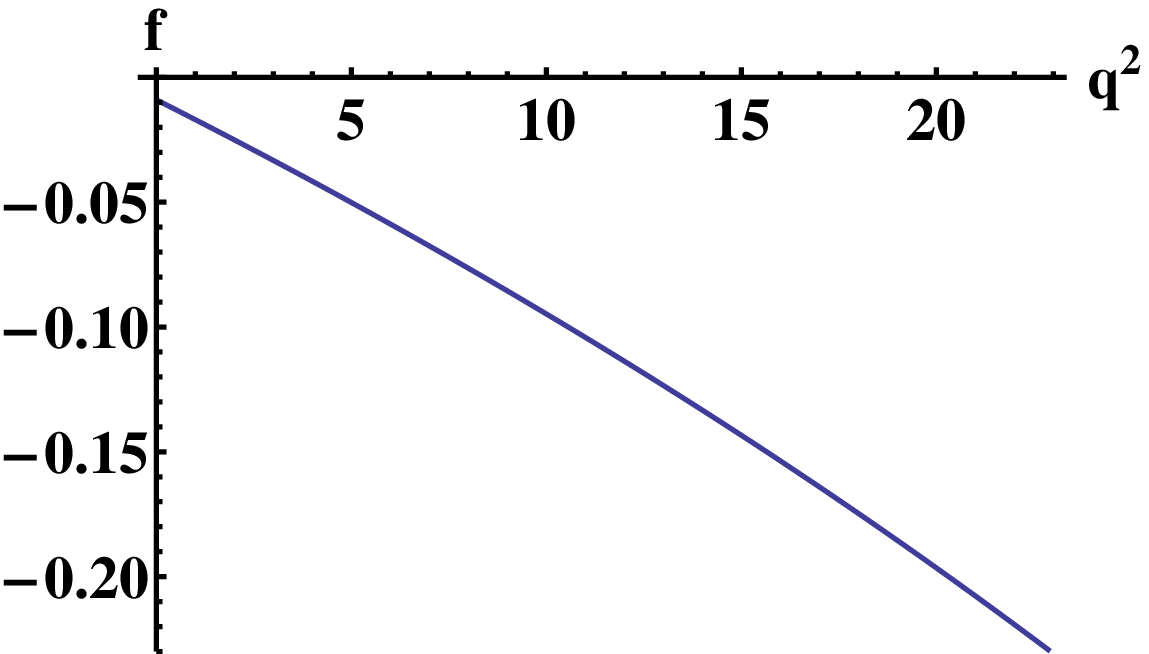}
\end{picture}    \hspace{20mm}
\begin{picture}(150,125)(-20,0)
  \includegraphics[height=.20\textheight]{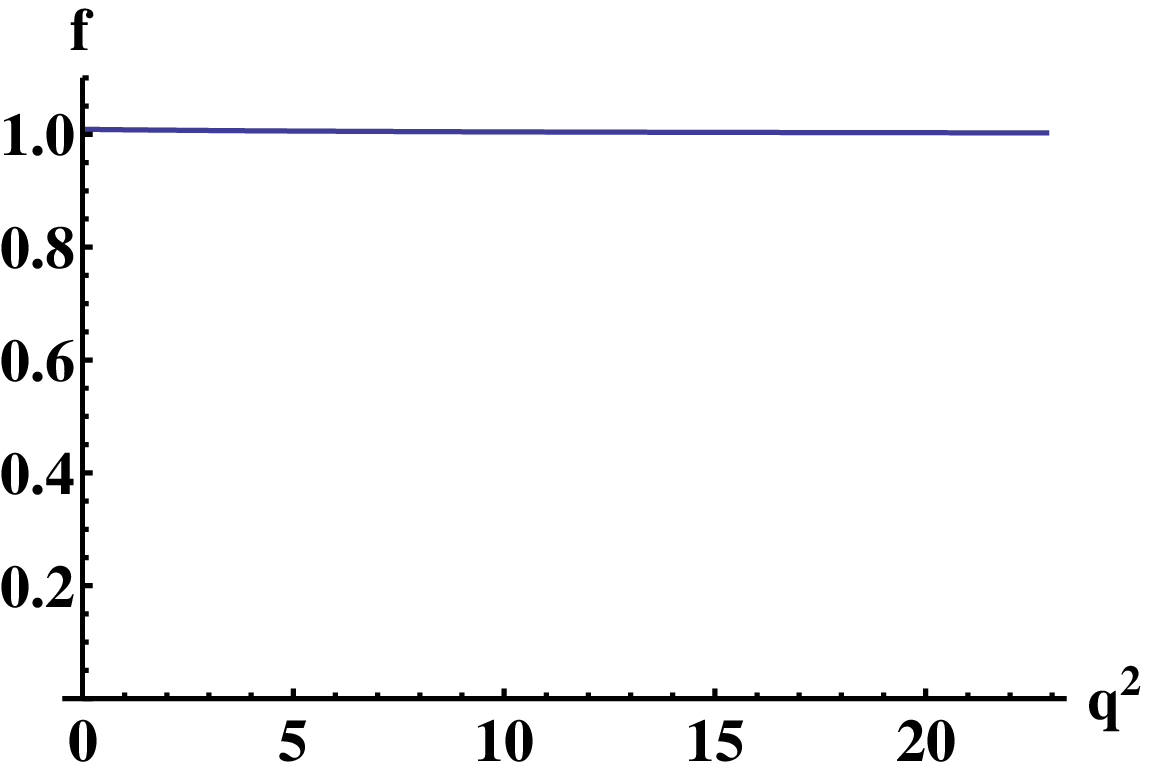}
\end{picture}

\hspace*{5mm} 
{(a) $f_+^a(q^2)$} \hspace{65mm} 
{(b) $f_+^b(q^2)$}

\begin{picture}(25,135)(110,0)
  \includegraphics[height=.20\textheight]{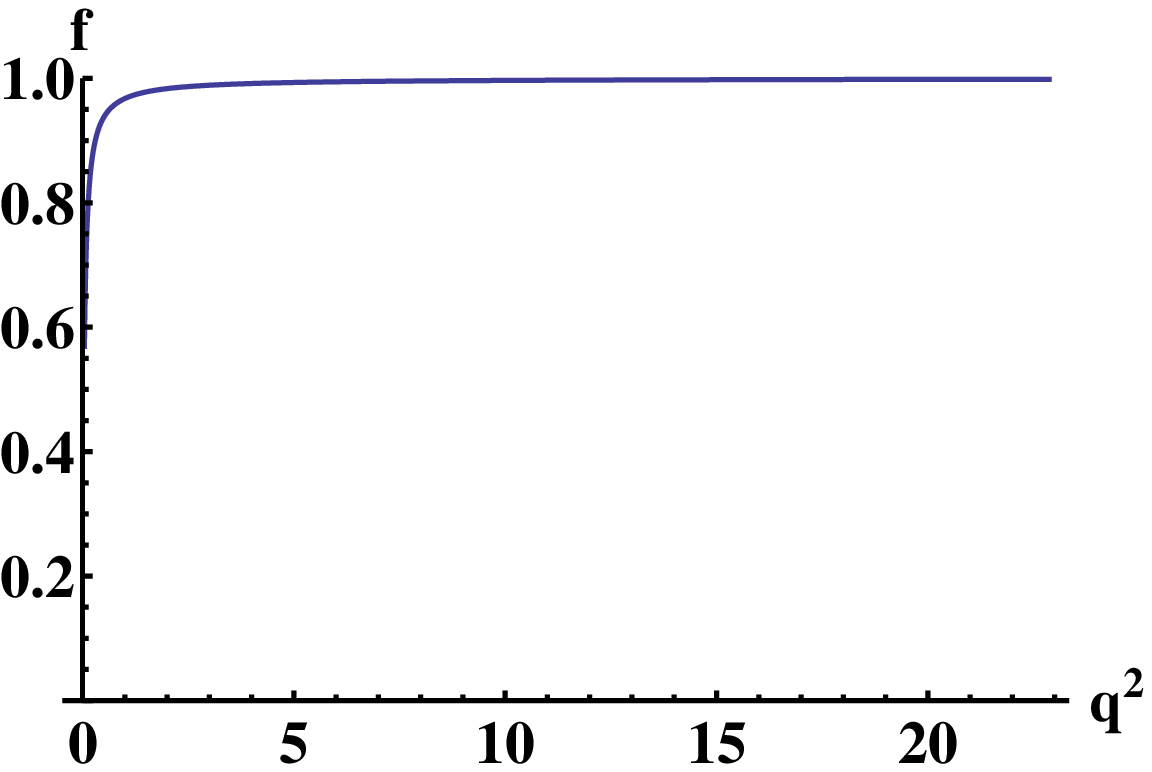}
\end{picture}    \hspace{20mm}
\begin{picture}(130,135)(-40,0)
  \includegraphics[height=.20\textheight]{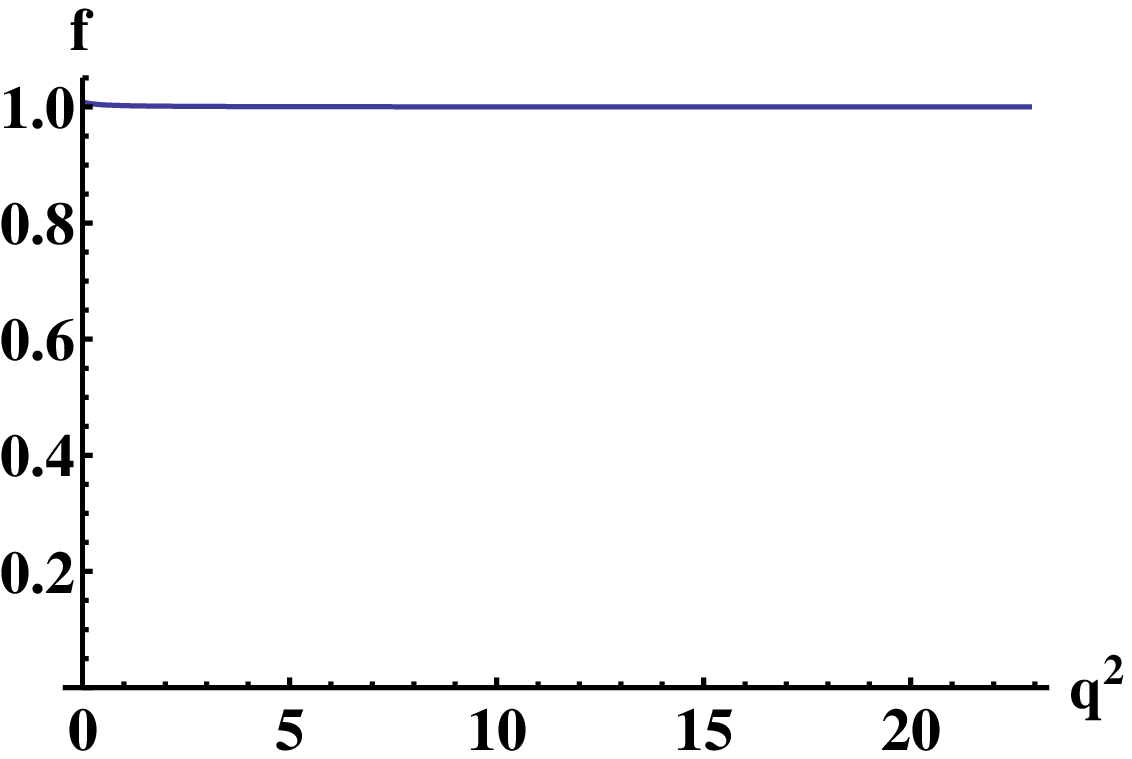}
\end{picture}

\hspace*{5mm} {(c) $f_+^c(q^2)$} \hspace{65mm} 
{(d) $f_+^d(q^2)$}
\end{center}
\caption{ 
Contribution from each quark line to the functions $f_+(q^2)$.  
Figures are illustrated for a physical range 
$4 m_\mu^2 <q^2 < (M_B -M_K)^2$. 
}
\label{Fig:4}
\end{figure}

\begin{figure}[!h]
\begin{center}
  \includegraphics[height=.25\textheight]{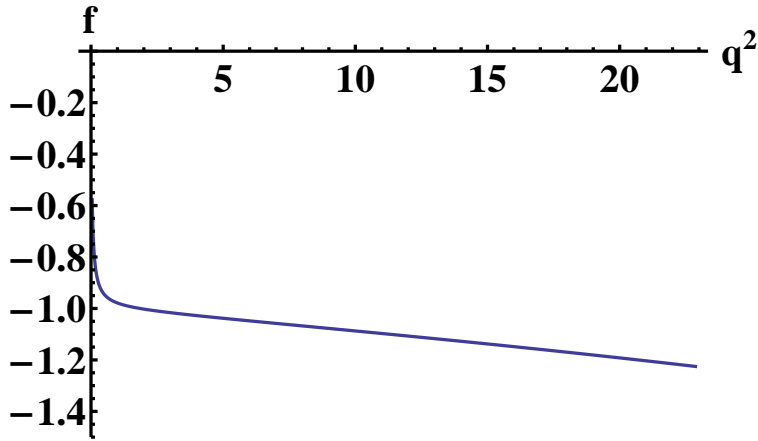}
\end{center}
\caption{Behavior of $f_+(q^2)$ in the neutral B meson decay 
$B^0 \rightarrow K^0 \ell^+ \ell^-$.  
 }
\label{Fig:5}
\end{figure}

First, in Fig.~\ref{Fig:4}, we show the behavior of the functions 
$f_+^a(q^2)$, $f_+^b(q^2)$, $f_+^c(q^2)$ and $f_+^d(q^2)$ 
which represent the contributions of photon emissions from 
$b$, $s$, $d_1$ and $d_2$ quarks, respectively, and which
are due to quark propagator effects. 
Note that although we have chosen the coefficients 
$(a_1, b_1)$ and $(a_2,b_2)$ so that those are independent 
of $q^2$, the functions $f_+^a(q^2)$,
$f_+^b(q^2)$, $f_+^c(q^2)$ and $f_+^d(q^2)$ still depend on $q^2$. 
We find that $f_+^b(q^2) \simeq +1$ and $f_+^d(q^2) \simeq +1$ 
for whole range of $q^2$, and $f_+^c(q^2) \simeq +1$
except for a small range of $q^2$.
Also, we show the behavior of $f_+(q^2)$ in Fig.~\ref{Fig:5}. 
Note that $f_+(q^2) < 0$ over the whole physical region. 

\begin{figure}[!h]
\begin{center}
  \includegraphics[height=.25\textheight]{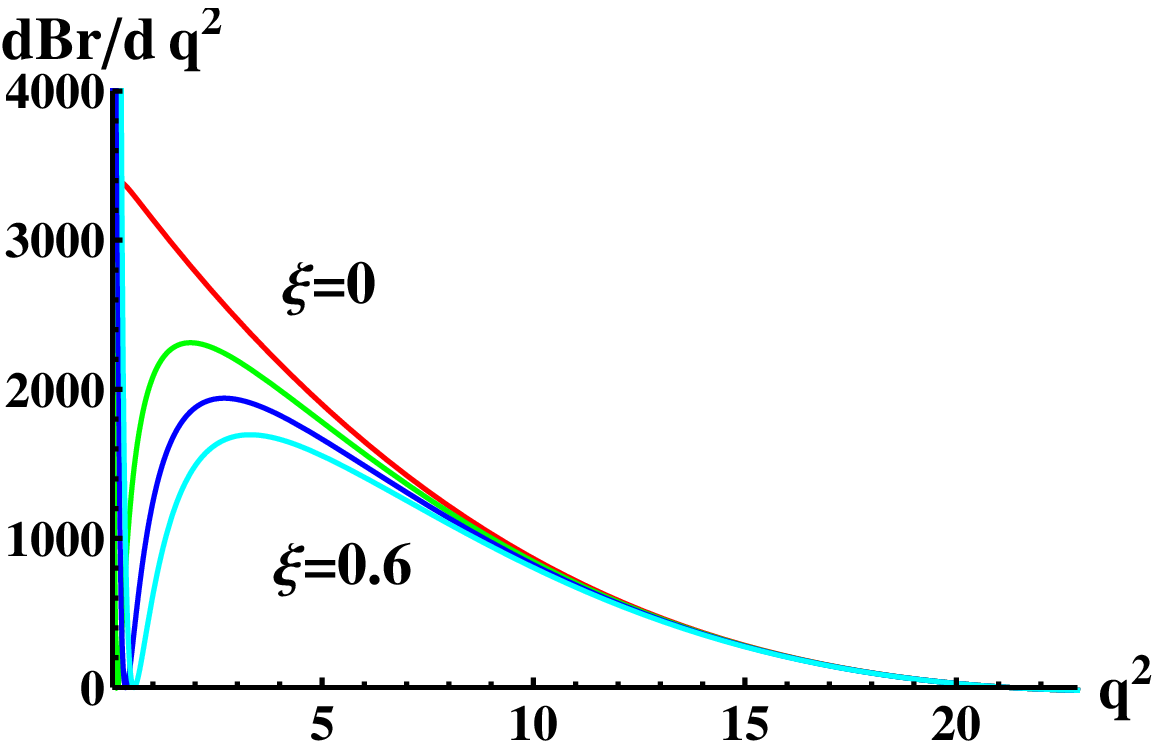}
\end{center}
\caption{Behavior of $d Br/d q^2$ in the decay 
$B^0 \rightarrow K^0 \ell^+ \ell^-$ 
in the unit of $G^2$ defined by Eq.~(\ref{Eq:5.2}).  
Curves are lined up in order of the cases $\xi=0$,  $0.2$, $0.4$ 
and $0.6$ in the unit of GeV$^2$ (the colors red, green, blue and 
cyan, respectively).
 }
\label{Fig:6}
\end{figure}

Also, we show the behavior of 
$d Br (B^0 \rightarrow K^0 \ell^+ \ell^-)/d q^2$ 
in the unit of $G^2$ defined by Eq.~(\ref{Eq:5.2}) 
for typical values of the parameter $\xi$ in Fig.~\ref{Fig:6}.
We can obtain a reasonable dip at $q^2 \sim 1$ GeV with $\xi = 0.6$.

\begin{figure}[!h]
\begin{center}
  \includegraphics[height=.25\textheight]{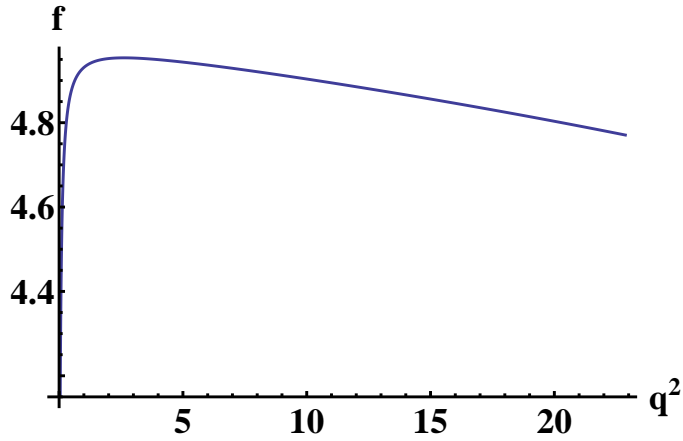}
\end{center}
\caption{
Behavior of $f_+(q^2)$ in the charged B meson decay 
$B^+ \rightarrow K^+ \ell^+ \ell^-$.  
}
\label{Fig:7}
\end{figure}

\begin{figure}[!h]
\begin{center}
  \includegraphics[height=.25\textheight]{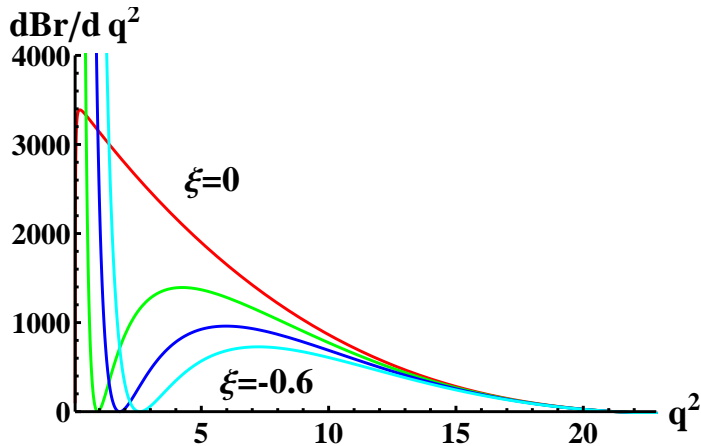}
\end{center}
\caption{
Behavior of $d Br/d q^2$ in the decay 
$B^+ \rightarrow K^+ \ell^+ \ell^-$
in the unit of $G^2$ defined by Eq.~(\ref{Eq:5.2}).  
Curves are lined up in order of the cases $\xi=0$,  $-0.2$, 
$-0.4$ and $-0.6$ in the unit of GeV$^2$ (the colors red, 
green, blue and cyan, respectively).
}
\label{Fig:8}
\end{figure}

Similarly, we can demonstrate the case of $B^+ \rightarrow K^+ \ell^+ \ell^-$.
The behaviors of $f_+(q^2)$ and  $d Br/d q^2$  are illustrated 
in Figs.~\ref{Fig:7} and \ref{Fig:8}, respectively. 
(Here, for convenience, we have used the same value of $G$ defined
by Eq.~(\ref{Eq:5.2}), although a weak annihilation diagram effect \cite{WA}
should be added in the case of $B^+ \rightarrow K^+ \ell^+ \ell^-$.) 
If the up-quark mixing is sizable compared with 
the down-quark mixing, the case will be also visible. 
The shape of the $d Br/d q^2$ in $B^+ \rightarrow K^+ \ell^+ \ell^-$
is almost similar to that in $B^0 \rightarrow K^0 \ell^+ \ell^-$. 
However, note that the dip in $d Br/d q^2$ appears for $\xi >0$
in the case  $B^0 \rightarrow K^0 \ell^+ \ell^-$, while the dip
appears for $\xi <0$ in the case $B^+ \rightarrow K^+ \ell^+ \ell^-$.
It will be possible because $U^{\ast u}_{21} U^u_{31}$ takes 
an opposite sign to $U^{\ast d}_{21} U^d_{31}$.
Moreover, the position of the dip is slightly shifted to the larger value of $q^2$ 
than the case of neutral B meson.

However, we do not consider that the magnitude of 
$U^{\ast u}_{21} U^u_{31}$ is accidentally the same 
as that of $U^{\ast d}_{21} U^d_{31}$.
We expect that the behavior of $d Br/d q^2$ 
in $B^+ \rightarrow K^+ \ell^+ \ell^-$ will be 
different from that in $B^0 \rightarrow K^0 \ell^+ \ell^-$.
We hope data of $d Br/d q^2$ will be able to distinguish 
between $B^0 \rightarrow K^0 \ell^+ \ell^-$ and
$B^+ \rightarrow K^+ \ell^+ \ell^-$.

\begin{figure}[!h]
\begin{center}
  \includegraphics[height=.55\textheight,angle=-90]{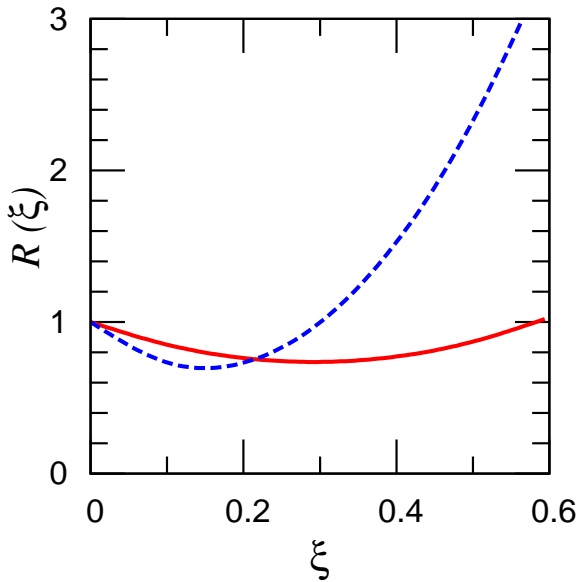}
\end{center}\vspace{-5mm}
\caption{
Behaviors of $R(\xi) \equiv \Gamma(\xi)/\Gamma(0)$ in the decays
$B^0 \rightarrow K^0 \ell^+ \ell^-$ (solid curve) and
$B^+ \rightarrow K^+ \ell^+ \ell^-$ (dashed curve).
}\label{Fig:9}
\end{figure}

Finally, we would like to give some comments on 
the predicted partial decay width.
The decay width is given by
\begin{eqnarray}
\Gamma (B \rightarrow K \ell^+ \ell^-) = G^2  
\int_{q^2_{min}}^{q^2_{max}} d q^2 F(q^2) \,,\label{Eq:6.3}
\end{eqnarray}
where the function $F(q^2)$ is defined by Eq.~(\ref{Eq:5.4})
and $q^2_{min}$ is given by $q^2_{min}= 4 m_\ell^2$.
The numerical value is highly sensitive to 
whether $\ell= \mu$ or $\ell= e$, because 
the contribution becomes very large at $q^2 \simeq 0$.
However, it seems to be impossible to measure 
accurately until $q^2= 4 m_e^2=1.044\times 10^{-6} $ GeV$^2$. 
If we take $q^2_{min}= 4 m_\mu^2=0.04465$ GeV$^2$ 
for the case of $\Gamma (B \rightarrow K e^+ e^-)$, too, 
we cannot find a significant difference between 
$\Gamma (B \rightarrow K e^+ e^-)$ and 
$\Gamma (B \rightarrow K \mu^+ \mu^-)$. 
Another comment is as follows:
The predicted decay width $\Gamma (B \rightarrow K \ell^+ \ell^-)$ 
is dependent on the value of $\xi$. 
We illustrate the behavior  $R(\xi) \equiv \Gamma(\xi)/\Gamma(0)$
in Fig.~\ref{Fig:9}. 
The present data \cite{PDG12} show 
$Br(B^+ \rightarrow K^+ \ell^+\ell^-)
= (5.1 \pm 0.5) \times 10^{-7}$, 
$Br(B^0 \rightarrow K^0 \ell^+\ell^-)
= (3.1^{+0.8}_{-0.7}) \times 10^{-7}$ and  
$\tau(B^+)/\tau(B^0)= 1.079 \pm 0.007$, 
so that we obtain
\begin{eqnarray}
R_{+/0} \equiv \frac{\Gamma(B^+ \rightarrow K^+ \ell^+\ell^-)}{
\Gamma(B^0 \rightarrow K^0 \ell^+\ell^-)} = 
1.52^{+0.42}_{-0.38}\,.\label{Eq:6.4}
\end{eqnarray}
Although the value has a large error, 
if we dare to take the center value in (\ref{Eq:5.3}),
a case of the value of $\xi$ which gives 
$R_{+/0} \sim 1.5$ is only in the case 
$B^+ \rightarrow K^+ \ell^+\ell^-$.
The case with $\xi \sim 0.4$ GeV$^2$ can also give
a reasonable shape of $d \Gamma/d q^2$ 
as seen in Fig.~\ref{Fig:8}. 
However, this view conflicts with our anticipation that 
$|U^{\ast u}_{21} U^u_{31}| \ll |U^{\ast d}_{21} U^d_{31}|$.
We must wait individual future data of 
$B^0 \rightarrow K^0 \ell^+ \ell^-$ and
$B^+ \rightarrow K^+ \ell^+ \ell^-$. 
%

%

\section{Concluding remarks}

In conclusion, we have investigated a contribution of
photon emission from the ``spectator'' quark $d \rightarrow d +\gamma$
($u \rightarrow u +\gamma$) in the $B^0$ ($B^+$) meson, and 
thereby we have obtained interesting 
results:
(i) The contribution from the spectator quark is 
not so negligible, i.e. $f_+^c(q^2) \simeq 1$ and $f_+^d(q^2) \simeq 1$
in contrast to the contribution from $\bar{b}$ quark, 
$f_+^a(q^2) \sim -0.1$, and that from $\bar{s}$ quark, 
$f_+^b(q^2) \simeq 1$.  
(ii) For a sizable value of the parameter $|\xi|$,
we can demonstrate a dip 
of $Br(B \rightarrow K \ell^+ \ell^-)$ in the small $q^2$ region.
However, in order to obtain such a dip in both decay modes,
$B^0 \rightarrow K^0 \ell^+ \ell^-$ and 
$B^+ \rightarrow K^+ \ell^+ \ell^-$, simultaneously,
the sign of $\xi$ parameter must be opposite each other.

However, note that 
it is hard to compare the expression (\ref{Eq:1.3}) 
with the prediction of the decay 
$B \rightarrow K\ell^+\ell^-$ from the standard model directly. 
Nevertheless the results $f_+(q^2)$ given in Fig.~\ref{Fig:4} 
are independent of 
the form $F(q^2)$ defined by Eq.~(\ref{Eq:5.4}).
Actually, the $q^2$ dependence of $d Br/dq^2$ is  correlated with 
the form $F(q^2)$, 
but in the present analysis, we have not taken QCD corrections, 
for example form factor effects, in order to demonstrate 
photon emission from the spectator quark straightforwardly.
Therefore, correspondingly to the treatments, we also 
simplify the conventional standard model contributions, too. 
The purpose of the present paper is to indicate 
the spectator quark effects qualitatively, and not
to estimate the spectator quark effects quantitatively.
In the numerical analysis, since our interest is 
the difference between $d Br(B^0 \rightarrow K^0 \ell^+ \ell^-)/d q^2$
and $d Br(B^+ \rightarrow K^+ \ell^+ \ell^-)/d q^2$, 
for simplicity, we have neglected some important effects.
For example, we have regarded the form factor $f_T(q^2)$ 
as a constant in respect to $q^2$.
Therefore, the numerical results should be rigidly taken.
However, we consider that the qualitative conclusions 
are reliable since we have treated only relative quantities 
({\it e.g.} the ratios). 

In the present paper, the origin of $b$-$s$ transition is not
specified, although, for convenience, the formulation has been 
given for the case of a family gauge boson $A_2^3$ exchange. 
In the present paper, the $b$-$s$ transition is given in the 
Eq.~(\ref{Eq:4.1}), but the definition (\ref{Eq:4.2}) of the coupling 
constant $G^q_{fam}$ is nothing but an example.
The parameter $\xi$ given in Eq.~(\ref{Eq:5.3}) is a phenomenological one, 
at present. 
The value of $\xi$ has been treated as one which should be determined by 
experiments. 

If we consider that the origin is due to the exchange of $A_2^3$ for example, 
the rough estimate of $|\xi|$ from Eq.~(\ref{Eq:5.3}) gives 
$|\xi| \sim 10^{-5}$ GeV$^2$ for $M_{23} \sim$ a few TeV. 
Accordingly, such contribution cannot become visible in the family gauge 
boson model even if it is inverted mass hierarchy 
\cite{K-Y_PLB_2012} (also in a revised model \cite{YK_PRD_2013}).
We need some enhancement mechanism of the $A_2^3$ exchange diagrams 
or some other dynamics in such rare B meson decay.\footnote{
A possibility that $A_2^3$ becomes considerably light is still not
ruled out.
As seen in \ref{App:A}, the observed value of $\Delta m(B_s)$ 
put on a constraint only for the mass $M_{22}$ (not for $M_{23}$).
Previously, we have speculated  $M_{23} \sim$ a few TeV 
\cite{YK_PRD_2013}
from a deviation from $e$-$\mu$ universality in the tau decays 
$\tau \rightarrow \mu \nu \bar{\nu}/e \nu \bar{\nu}$.
However, the value was obtained by assuming $M_{23} \ll M_{13}$.
If $M_{23} \simeq M_{13}$, we cannot extract a value of $M_{23}$
from the decays $\tau \rightarrow \mu \nu \bar{\nu}/e \nu \bar{\nu}$.  
A possibility that $M_{33}, M_{23} \sim 10^{-1}$ TeV is still 
not ruled out. }

On the other hand, we have other diagrams for the source of 
$b$-$s$ transition, electroweak penguin, gluon penguin, 
and other considerable processes.
Especially, so far, the gluon penguin has been neglected
in the operator expansion approach. 
If we replace the family gauge boson $A_2^3$ with 
gluon $g$ from the gluon penguin, the value of $\xi$ 
can be sizable.
Therefore, we may rather regard the parameter $\xi$ defined 
by Eq.~(\ref{Eq:5.1})
as a phenomenological one, discarding Eq.~(\ref{Eq:5.3}).
Then, the squared mass $M_{23}^2$ in Eq.~(\ref{Eq:4.2}) must be replaced with 
$\bar{q}^2 =(\bar{p}_1 -\bar{p}_2)^2$. 
The value $\bar{q}^2$ is calculable in the 
present prescription. 
Since our parameters $a_1$, $b_1$, $a_2$ and $b_2$ are small,
the value $\bar{q}^2$ is the order of $q^2$.
Therefore, the $q^2$ dependence will be somewhat 
different from the present result based on the
$A_2^3$ exchange.

In the case of gluon penguin, the decay widths of $B^0$ and $B^+$ 
decays are given by the same forms except for 
the factors $f_+(q^2)$.
Since the parameters $\xi(B^0)$ and $\xi(B^+)$ are also 
given by the same value,
the dip in $d \Gamma/d q^2$ can appear only in either 
$B^0$ or $B^+$ decay.
(For the case of $A_2^3$ exchange, 
$\xi(B^0)$ and $\xi(B^+)$ can take 
opposite sign each other by supposing 
$U^{\ast u}_{21} U^u_{31}/U^{\ast d}_{21} U^d_{31} <0$.)
At present, the data by Belle \cite{Belle} and BABAR \cite{Babar} 
have shown a possibility that there is a dip in   $d \Gamma/d q^2$,
but data are not separated between $B^0$ and $B^+$. 
In the LHCb, we can see a possibility of a dip 
in the $B^0$ decay \cite{LHCb2012dip}, but we cannot see such a dip in the
$B^+$ decay \cite{LHCb2012non}. 
It seems that this is favor of the gluon penguin model. 

Thus, it is our greatest concern whether the data 
show a dip in $d Br/d q^2$ both or either in $B^0$
and/or $B^+$ decays. 
As in Fig.~\ref{Fig:9}, we can find a possibility of appearance 
for isospin asymmetry, 
that is the difference between $B^0$ and $B^+$.
In the evaluation of that figure, 
the overall factors should be canceled out
because we have taken a ratio of the decay rate.
Therefore the phenomenon of photon emission from spectator quarks itself is 
important to observe isospin asymmetry.
We expect that such data will soon be reported.

The present results highly depend on our treatment for 
the quark-anti-quark bound system.
In our prescription, the existence of the quark propagator,
which cannot be incorporated into the factorization method, 
has played an essential role.
We have straightforwardly and faithfully calculated 
the effects based on the effective valence quark model.
We think that the present prescription should be worthwhile 
to be tested by future experimental data.



\section*{Acknowledgments}
The authors thank M.~Tanaka and Y.~Okada for helpful suggestions
on this topic, 
and S.~Nishida and K.~Hayasaka for useful comments on the rare $B$ 
decay experiments. 
The authors also thank T.~Feldmann, A.~Khodjamirian and 
R.~Zwicky for helpful comments and informing valuable references.

%
\renewcommand\theequation{\thesection.\arabic{equation}}
\appendix
%

\def\thesection{Appendix.\Alph{section}}
\section{}\label{App:A}
\def\thesection{\Alph{section}}

In the present family gauge boson model \cite{K-Y_PLB_2012}, 
the family number changing interactions are exactly forbidden 
in the limit of absence of the quark mixings $U^u = {\bf 1}$ 
and $U^d = {\bf 1}$. 
In this Appendix, we give a brief review this family gauge boson
model.

The family gauge boson masses are generated \cite{Sumino_PLB09}
by a scalar $\Phi_{i\alpha}$
of $({\bf 3}, {\bf 3}^*)$ of U(3)$\times$U(3)$'$ which are broken 
at $\mu = \Lambda$ and $\mu=\Lambda'$ ($\Lambda \ll \Lambda'$),
respectively.
In the model, scalars  $({\bf 3}, {\bf 1})$ 
and  $({\bf 6}, {\bf 1})$ are absent. 

From the interactions (\ref{Eq:2.1}), 
effective interactions with $\Delta N_F=2$ are given as follows:
\begin{eqnarray}
H^{eff}_{\Delta N_F=2}  = \frac{g_F^2}{2}  \left[\sum_{i} 
\frac{\lambda_i^2}{M_{ii}^2} +
2 \sum_{i<j} \frac{\lambda_i \lambda_j}{M_{ij}^2} \right]
(\bar{q}_{k} \gamma_\mu q_{l}) (\bar{q}_{k}\gamma^\mu q_{l})
\equiv G_{eff} (\bar{q}_{k} \gamma_\mu q_{l}) 
(\bar{q}_{k}\gamma^\mu q_{l})\,.\label{Eq:App.A.1}
\end{eqnarray}
where 
\begin{eqnarray}
\lambda_i^{q} = U^{q*}_{ik} U^q_{il}\,,\label{Eq:App.A.2}
\end{eqnarray}
and, for simplicity, we have assumed $U_L^q =U_R^q$.
Note that, from the so-called unitary triangle, 
$\lambda_i$ satisfy 
\begin{eqnarray}
\lambda_1 + \lambda_2 + \lambda_3 =0\,.\label{Eq:App.A.3}
\end{eqnarray}

For convenience, let us take $U^d =V_{CKM}$. 
Then, for example, explicit values of $\lambda_i$ are
given as follows \cite{PDG12}:
\begin{eqnarray}
\lambda_1^2 \eqn{=} |V_{11}^* V_{12}|^2 = 4.81987 \times 10^{-2}\,, \ \ \ 
\lambda_2^2 = |V_{21}^* V_{22}|^2 = 4.80568 \times 10^{-2}\,, \notag\\
\lambda_3^2 \eqn{=} |V_{31}^* V_{32}|^2 = 1.2269 \times 10^{-7}
\,,\label{Eq:App.A.4}   
\end{eqnarray}
for $K^0$-$\bar{K}^0$ mixing, and
\begin{eqnarray}
\lambda_1^2 \eqn{=} |V_{12}^* V_{13}|^2 = 6.2559 \times 10^{-7}\,, \ \ \ 
\lambda_2^2 = |V_{22}^* V_{23}|^2 = 1.6085 \times 10^{-3}\,,\notag\\ 
\lambda_3^2 \eqn{=} |V_{32}^* V_{33}|^2 = 1.6294 \times 10^{-3}
\,,\label{Eq:App.A.5} 
\end{eqnarray}
for $B_s^0$-$\bar{B}_s^0$ mixing.
We can approximately regard $\lambda_i$ as $\lambda_3 \simeq 0$ and
$\lambda_1\simeq  -\lambda_2$  for $K^0$-$\bar{K}^0$ mixing, and
as $\lambda_1 \simeq 0$ and
$\lambda_3\simeq  -\lambda_2$  for $B_s^0$-$\bar{B}_s^0$ mixing.
Therefore, we can approximately express the effective 
coupling constant $G_{eff}$ as
\begin{eqnarray}
G_{eff}^K \simeq \frac{g_F^2}{2} \lambda_2^2 \left(
\frac{1}{M_{11}^2} +\frac{1}{M_{22}^2} -\frac{2}{M_{12}^2} \right) 
\simeq \frac{\lambda_2^2}{M_{22}^2}
\,,\label{Eq:App.A.6}
\end{eqnarray}
for  $K^0$-$\bar{K}^0$ mixing, and
\begin{eqnarray}
G_{eff}^{B_s} \simeq \frac{g_F^2}{2} \lambda_2^2 \left(
\frac{1}{M_{22}^2} +\frac{1}{M_{33}^2} -\frac{2}{M_{23}^2} \right) 
\simeq \frac{\lambda_2^2}{M_{33}^2}
\,,\label{Eq:App.A.7}
\end{eqnarray}
for $B_s^0$-$\bar{B}_s^0$ mixing.
Here, we have used guage boson mass relations 
$2 M_{ij}^2 = M_{ii}^2 + M_{jj}^2$ and an inverted mass hierarchy model
$M_{33}^2 \ll M_{22}^2 \ll M_{11}^2$  \cite{K-Y_PLB_2012}.
As seen in (A.7), as far as we do not consider too small mass value 
of $M_{33}$ (e.g. $\sim 10^2$ GeV), the model does not 
give a major contribution to the $B_s^0$-$\bar{B}_s^0$ mixing
$\Delta m_B^{obs}  =(1.164 \pm 0.005) \times 10^{-13}$ TeV \cite{PDG12}.  
This is independent of an explicit mass value of $M_{23}$. 
Note that, differently from the $\Delta N_F =2$ process in which 
a kind of the Glashow-Iliopoulos-Maiani mechanism \cite{GIM_PRD70} works, 
such a suppression does not work in the $\Delta N_F =1$
process in $B \rightarrow K$. 

Also, note that if we suppose our gauge boson masses are almost degenerated, 
the effective coupling constant $G_{eff}$ becomes nearly zero independently
of the mixing parameters $\lambda_i$, as seen from Eq.(A.1). 

Anyhow, in this model, we have a possibility that a value of
$M_{23}$ is considerably small.


\def\thesection{Appendix.\Alph{section}}
\section{}\label{App:B}
\def\thesection{\Alph{section}}

First, at quark level, 
we obtain the following amplitudes  
which correspond 
to the diagrams (a), (b), (c) and (d) in Fig.~\ref{Fig:3}:
\begin{eqnarray}
\hspace{-3mm}
{\cal M}^{eff}_{(a)} \eqn{=} i\frac{1}{6} \bar{e}_b e \left[ 
\bar{u}_d (p_2)\, \Gamma\, v_s (\bar{p}_2)
\right]\, \left[ \bar{v}_b (\bar{p}_1) \, \gamma_\mu \, 
\frac{\not\!\ell_{(a)} +m_b}{\ell_{(a)}^2 -m_b^2} \,
\Gamma \, u_d(p_1) \right] 
\frac{1}{q^2} [\bar{v}_\ell(k_2) \gamma^\mu u_\ell(k_1)] \,\label{Eq:App.B.1}\\
\hspace{-3mm}
{\cal M}^{eff}_{(b)} \eqn{=} i\frac{1}{6} \bar{e}_b e \left[ 
\bar{u}_d (p_2)\, \Gamma\, 
\frac{\not\!\ell_{(b)} +m_s}{\ell_{(b)}^2 -m_s^2} \,
\gamma_\mu \, v_s (\bar{p}_2)
\right]\, \left[ \bar{v}_b (\bar{p}_1) \, 
\Gamma \, u_d(p_1) \right] 
\frac{1}{q^2} [\bar{v}_\ell(k_2) \gamma^\mu u_\ell(k_1)] \,,\label{Eq:App.B.2}\\
\hspace{-3mm}
{\cal M}^{eff}_{(c)} \eqn{=}  i\frac{1}{6} e_d e \left[ \bar{u}_d (p_2)\, \Gamma\, 
v_s (\bar{p}_2)
\right]\, \left[ \bar{v}_b (\bar{p}_1) \, \Gamma \, 
\frac{\not\!\ell_{(c)} +m_d}{\ell_{(c)}^2 -m_d^2} \,
\gamma_\mu \, u_d(p_1) \right] 
\frac{1}{q^2} [\bar{v}_\ell(k_2) \gamma^\mu u_\ell(k_1)] \,,\label{Eq:App.B.3}\\
\hspace{-3mm}
{\cal M}^{eff}_{(d)} \eqn{=} i\frac{1}{6} e_d e \left[ \bar{u}_d (p_2)\, 
\gamma_\mu \, 
\frac{\not\!\ell_{(d)} +m_d}{\ell_{(d)}^2 -m_d^2} \,
 \Gamma\, v_s (\bar{p}_2)
\right]\, \left[ \bar{v}_b (\bar{p}_1) \, 
\Gamma \, u_d(p_1) \right] 
\frac{1}{q^2} [\bar{v}_\ell(k_2) \gamma^\mu u_\ell(k_1)] \,,\label{Eq:App.B.4}
\end{eqnarray}
where 
\begin{eqnarray}
\ell_{(a)} = \bar{p}_1 - q\,, \ \ \  
\ell_{(b)} = \bar{p}_2 + q\,, \ \ \   
\ell_{(c)} = {p}_1 - q\,, \ \ \  
\ell_{(d)} = {p}_2 + q\,, \label{Eq:App.B.5}
\end{eqnarray}
and the common coefficient $G_{fam}^{eff}$ has been 
dropped. 
Here, in order to provide for the next step in which
we obtain hadronic current form from the quark current 
form, the expressions (\ref{Eq:App.B.1}) - (\ref{Eq:App.B.4}) have been given by 
using a Fierz transformation 
\begin{eqnarray}
(\bar{b} \gamma_\rho s)(\bar{d}\gamma^\rho d) \ 
\Rightarrow \ 
\sum_{\Gamma} \left[ -\frac{1}{3} (\bar{d}\, \Gamma\, s)(
\bar{b}\, \Gamma\, d) -\frac{1}{2} \sum_{a=1}^{8} 
 (\bar{d}\, \Gamma \lambda_a\,  s)(\bar{b}\, \Gamma \lambda_a\, d)
\right] \,,\label{Eq:App.B.6}
\end{eqnarray}
where
\begin{eqnarray}
\Gamma \otimes \Gamma = - {\bf 1}\otimes {\bf 1} +
\gamma_5 \otimes \gamma_5  
+\frac{1}{2} \gamma_\rho \otimes \gamma^\rho 
+\frac{1}{2} \gamma_\rho \gamma_5 \otimes \gamma^\rho \gamma_5\,.\label{Eq:App.B.7}
\end{eqnarray}

Next, we must translate the amplitudes (\ref{Eq:App.B.1}) - (\ref{Eq:App.B.4}) 
in quark level into those in hadronic level.
We use the prescription (\ref{Eq:4.11}).
We obtain the following decay amplitudes from (\ref{Eq:App.B.1}) - (\ref{Eq:App.B.4}):
\begin{align}
{\cal M}_{a} &= i \frac{e^2}{18} f_K f_B 
\frac{1}{\Delta_a} \left[ (\bar{p}_1 -q)_\mu (P_B P_K)
+ P_{B\mu} (\bar{p}_1 -q)P_K - P_{K\mu} (\bar{p}_1 -q)P_B \right]
\frac{1}{q^2} [\bar{v}_\ell(k_2) \gamma^\mu u_\ell(k_1)] \,,\label{Eq:App.B.8}\\
{\cal M}_{b} &= i \frac{e^2}{18} f_K f_B 
\frac{1}{\Delta_b} \left[ (\bar{p}_2 +q)_\mu (P_B P_K)
+ P_{K\mu} (\bar{p}_2 +q)P_B - P_{B\mu} (\bar{p}_2 +q)P_K \right]
\frac{1}{q^2} [\bar{v}_\ell(k_2) \gamma^\mu u_\ell(k_1)] \,,\label{Eq:App.B.9}\\
{\cal M}_{c} &= -i \frac{e^2}{18} f_K f_B 
\frac{1}{\Delta_c} \left[ (p_1-q)_\mu (P_B P_K)
+ P_{B\mu} ({p}_1 -q)P_K - P_{K\mu} ({p}_1 -q)P_B \right]
\frac{1}{q^2} [\bar{v}_\ell(k_2) \gamma^\mu u_\ell(k_1)] \,,\label{Eq:App.B.10}\\
{\cal M}_{d} &= -i \frac{e^2}{18} f_K f_B 
\frac{1}{\Delta_d} 
\left[ (p_2 +q)_\mu (P_B P_K)
+ P_{K\mu} ({p}_2 +q)P_B - P_{B\mu} ({p}_2 +q)P_K \right]
\frac{1}{q^2} [\bar{v}_\ell(k_2) \gamma^\mu u_\ell(k_1)] \,,\label{Eq:App.B.11}
\end{align}

When we use the expression (\ref{Eq:3.4}), we obtain the following
form for the meson currents: 
\begin{align}
{\cal M} = i \frac{e^2}{18} f_K f_B \frac{1}{2}
\left[ f_+ (q^2) (P_B +P_K)_\mu + f_- (q^2) (P_B -P_K)_\mu \right]
\frac{1}{q^2} [\bar{v}_\ell(k_2) \gamma^\mu u_\ell(k_1)] \,.\label{Eq:App.B.12}
\end{align}
The second term with $q_\mu =(P_B -P_K)_\mu$ in Eq.~(\ref{Eq:App.B.12}) does not 
contribute the decay amplitude because of
$q_\mu [\bar{v}_\ell(k_2) \gamma^\mu u_\ell(k_1)] =0$ for 
$m_{\ell 1} = m_{\ell 2}$.
For the expression $f_+(q^2)$, we obtain
\begin{align}
f_+(q^2) = f_{+}^a(q^2) + f_{+}^b(q^2) - f_{+}^c(q^2) - f_{+}^d(q^2) \,,\label{Eq:App.B.13}
\end{align}
where
\begin{eqnarray}
f_{+}^a(q^2) \eqn{=} \frac{ (x_1-2 a_1)M_K^2 +(1-x_1+ a_1 +b_1) q^2 }{
-(x_1-2 a_1) \Delta_{BK}^2 +(1-x_1+ 2 b_1) q^2}\,,\label{Eq:App.B.14}\\
f_{+}^b(q^2) \eqn{=} \frac{ (x_2-2 a_2) M_K^2 +(1-x_2+ a_2 +b_2) q^2 }{
(x_2-2 a_2) \Delta_{BK}^2 +(1-x_2+ 2 b_2) q^2}\,,\label{Eq:App.B.15}\\
f_{+}^c(q^2) \eqn{=} \frac{ 2 a_1 M_K^2 + ( 1-a_1 -b_1) q^2 }{ 
 - 2a_1 \Delta_{BK}^2  + (1 -2 b_1) q^2} \,,\label{Eq:App.B.16}\\
f_{+}^d(q^2) \eqn{=} \frac{ 2 a_2 M_B^2 + (1- a_2 -b_2) q^2}{
2a_2 \Delta_{BK}^2  + (1 -2 b_2) q^2} \,.\label{Eq:App.B.17}
\end{eqnarray}

A form of $f_+(q^2)$ for the decay $B^+ \rightarrow K^+ \ell^+ \ell^-$ can
be obtained by replacing $e_d=-e/3 \rightarrow e_u=+2e/3$ in (\ref{Eq:App.B.13}): 
\begin{align}
f_+(q^2) = f_{+}^a(q^2) +f_{+}^b(q^2) +2 f_{+}^c(q^2)+2 f_{+}^d(q^2) \,.\label{Eq:App.B.18}
\end{align}

%
\def\thesection{Appendix.\Alph{section}}
\section{}\label{App:C}
\def\thesection{\Alph{section}}

The function $F(q^2)$ corresponds to $d \Gamma/d q^2$ for the conventional 
electroweak photon penguin, and it is calculated from the matrix element
\begin{align}
{\cal M} = G (P_B+P_K)_\mu  \bar{\ell}(k_2) \gamma^\mu \ell(k_1) \,,\label{Eq:App.C.1}
\end{align}
where $G$ is defined by Eq.~(\ref{Eq:5.2}).
By defining a parameter $ y \equiv m^2_{\ell K} =(k_2 +P_K)^2$ 
together with $y_1 = y_{min}$ and $y_2=y_{max}$, 
the form $F(q^2)$ is represented as
\begin{align}
G^2 F(x) &\equiv \frac{1}{(2\pi)^3} \frac{1}{32 M_B^3} 
\int_{y_1}^{y_2} dy\, |{\cal M}|^2\notag \\
&=- \frac{1}{(2\pi)^3} \frac{1}{32 M_B^3} \left[
\frac{1}{3} (y_2^3 -y_1^3) +\frac{1}{2} a (y_2^2 -y_1^2) + b (y_2 -y_1) 
\right] \,,\label{Eq:App.C.2}
\end{align}
where 
\begin{align}
a&=q^2-(M_B^2 + M_K^2 + 2m_\ell^2) \,, \\
b&=(M_B^2+M_K^2)(M_K^2+m_\ell^2) - m_\ell^2 q^2 \,.\label{Eq:App.C.3}
\end{align}

\def\thesection{Appendix.\Alph{section}}
\section{}\label{App:D}
\def\thesection{\Alph{section}}

The coefficients $(a_1, b_1)$ can be obtained as follows.
When we define 
\begin{eqnarray}
A \eqn{=} 2 (M_B^2+M_K^2) - q^2 , \ \ \ B = q^2, \ \ \ C = \Delta_{BK}^2\,,\label{Eq:App.D.1}
\end{eqnarray}
from Eq.~(\ref{Eq:3.5}), we obtain a relation between $a_1$ and $b_1$: 
\begin{eqnarray}
b_1 \eqn{=} \frac{1}{B} \left[ -C a_1 \pm \sqrt{ D a_1^2 + B m_{d1}^2} 
\right] \,,\label{Eq:App.D.2}
\end{eqnarray}
i.e.
\begin{eqnarray}
b_1 \eqn{=} \frac{1}{q^2} \left[ - a_1 \Delta_{BK}^2 
\pm \sqrt{ D a_1^2 + m_{d1}^2 q^2} 
\right] \,,\label{Eq:App.D.3}
\end{eqnarray}
where
\begin{eqnarray}
D \eqn{\equiv} C^2 -AB = 
(\Delta_{BK}^2 )^2 - q^2 [ 2(M_B^2 + M_K^2) -q^2 ]\notag\\
\eqn{=} \left[ (M_B -M_K)^2 -q^2\right] \left[ (M_B +M_K)^2 -q^2\right] \,.\label{Eq:App.D.4}
\end{eqnarray}
By substituting Eq.~(\ref{Eq:App.D.4}) into Eq.~(\ref{Eq:3.7}), we obtain a 
relation for $a_1$
\begin{eqnarray}
x_1^2 M_B^2 + m_{d1}^2 -m_b^2 = \frac{x_1}{q^2} \left[ - a_1 D 
\pm (\Delta_{BK}^2 +q^2 ) \sqrt{a_1^2 D + m_{d1}^2 q^2} 
\right] \,.\label{Eq:App.D.5}
\end{eqnarray}
The parameter $a_1$ can be obtained by solving Eq.~(\ref{Eq:App.D.5}) 
for $a_1$.

\def\thesection{Appendix.\Alph{section}}
\section{}\label{App:E}
\def\thesection{\Alph{section}}

More exactly speaking, the Eq.~(\ref{Eq:5.1}) should be replaced by
\begin{eqnarray}
{\cal M} = G(q^2) \left( 1 + \xi \frac{f_T(0)}{f_T(q^2)}  \frac{f_+(q^2)}{q^2} \right) 
(P_B +P_K)_\mu [\bar{v}_\ell(k_2) \gamma^\mu u_\ell (k_1) ] \,,\label{Eq:App.E.1}
\end{eqnarray}
where 
\begin{eqnarray}
G(q^2) = G^{eff}_{EW} \frac{2 m_b f_T(q^2)}{M_B + M_K} \,.\label{Eq:App.E.2}
\end{eqnarray}
The parameter $\xi$ is defined by
\begin{eqnarray}
\xi =  \frac{g_{fam}^2}{g_w^2} \frac{8 M_w^2}{M_{23}^2} 
\frac{U_{33}^{\ast d} U_{22}^d U_{21}^{\ast d} U_{31}^d }{V_{ts}^* V_{tb} } 
\frac{\pi^2}{9} 
 \frac{M_B + M_K}{2 m_b f_T(0)} f_K f_B \,,\label{Eq:App.E.3}
\end{eqnarray} 
which is unchanged from Eq.~(\ref{Eq:5.3}).
Then, $d \Gamma/d q^2$ is given by
\begin{eqnarray}
\frac{d\Gamma}{d q^2} (B \rightarrow K \ell^+ \ell^-) =  
G^2(q^2) \left( 1 + \xi\frac{f_T(0)}{f_T(q^2)}  
\frac{f_+(q^2)}{q^2} \right)^2 F(q^2) \,.\label{Eq:App.E.4}
\end{eqnarray}

\begin{figure}[!h]
\begin{center}
  \includegraphics[height=.25\textheight]{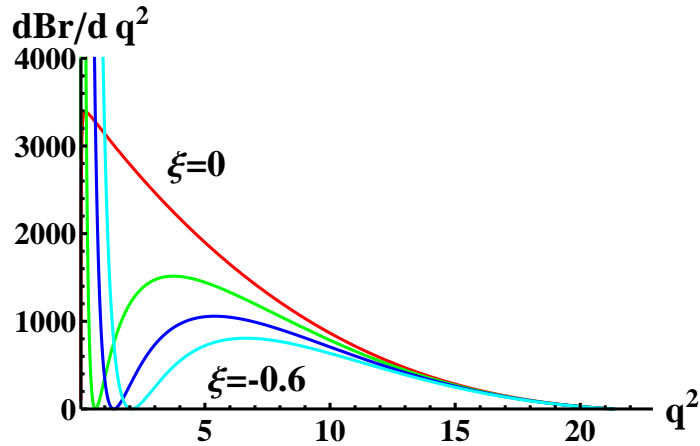}
\end{center}
\caption{Behavior of $d Br/d q^2$ in the decay 
$B^0 \rightarrow K^0 \ell^+ \ell^-$ 
in the unit of $G^2$ defined by Eq.~(\ref{Eq:App.E.2}).  
Curves are lined up in order of the cases $\xi=0$,  $0.2$, $0.4$ 
and $0.6$ in the unit of GeV$^2$ (the colors red, green, blue and 
cyan, respectively).
 }\label{Fig:10}
\end{figure}

In order to compare with the over-simplified previous result
Fig.~\ref{Fig:6}, we illustrate the behavior of $d Br/dq^2$ for 
the same value of $\xi$, 
where parameters of the form factor $f_T(q^2)$ have 
been quoted from Ref.\cite{form_factor}. 
We can see that the numerical results are almost not changed 
between Fig.~\ref{Fig:6} and Fig.~\ref{Fig:10}. 

\vspace{10mm}

%
%

\end{document}